\documentclass[prd,twocolumn,a4paper,showpacs]{revtex4}

\usepackage{hyperref}

\usepackage{graphicx}
\usepackage{subfigure}

\usepackage{amsmath}
\usepackage{mathtools}
\usepackage{MnSymbol}
\usepackage{dsfont}
\usepackage{bbold}
\usepackage{slashed}

\usepackage[squaren]{SIunits}

\DeclareMathOperator\arctanh{arctanh}
\newcommand{\tr}[1]{\operatorname{tr} \left\{ #1 \right\}}
\newcommand{\Dr}{ \overset{\rightarrow}{D} }
\declareslashed{}{/}{.07}{-.3}{\Dr}
\newcommand{\Dl}{ \overset{\leftarrow}{D} }
\declareslashed{}{/}{.07}{-.3}{\Dl}

\newcommand{\vshiftindex}[2]{\mathrel{\raisebox{#2}{$_#1$}}}

\begin{document}
\allowdisplaybreaks
\begin{abstract}
In this paper we present chiral extrapolation formulas for the generalized form factors connected to the first moments of chiral-even generalized parton distributions. For the description of QCD at low energies we employ the framework of two-flavor covariant baryon chiral perturbation theory at full one-loop accuracy. Our results are well suited for the extrapolation of lattice QCD data to physical quark masses.
\end{abstract}

\title{First moments of nucleon generalized parton distributions in chiral perturbation theory at full one-loop order}
\author{Philipp \surname{Wein}}
\email[]{philipp.wein@physik.uni-regensburg.de}
\author{Peter C. \surname{Bruns}}
\email[]{peter.bruns@physik.uni-regensburg.de}
\author{Andreas \surname{Sch\"afer}}
\email[]{andreas.schaefer@physik.uni-regensburg.de}
\affiliation{Universit\"at Regensburg}
\maketitle

%
%
\section{Introduction}

By the end of the last century generalized parton distributions (GPDs) had been identified as a suitable tool to parametrize exclusive processes and to extract the contained information on hadron structure \cite{Mueller:1998fv,Ji:1996ek,Ji:1997gm,Radyushkin:1997ki} (for a more detailed description we refer to the reviews \cite{Diehl:2003ny,Ji:2004gf,Belitsky:2005qn}). This has triggered large activities in both the experimental and theoretical communities. On the experimental side one faces the difficulty that (hard) exclusive processes are power suppressed. Hence, precision measurements are challenging and require high luminosities. In addition, GPDs enter the cross section not directly, but via a convolution with a hard scattering kernel. Complementary information from the theoretical side is therefore most valuable. One framework capable of providing such information on nonperturbative quantities is lattice QCD. Unfortunately simulations suffer typically from at least one of the systematic errors associated with discretization, finite volume and unphysical quark masses. Therefore, having a better control over some of the various extrapolations and the associated systematic errors is very helpful to obtain reliable quantitative results. The most rigorous way to gain such control is given by chiral perturbation theory \cite{Weinberg:1978kz,Gasser:1983yg,Gasser:1987rb} as opposed to \emph{ad hoc} or model-based extrapolation formulas.\par
In this paper we focus on the first $x$ moments of chiral-even GPDs. In the parity-even sector those encode information on the momentum distribution of the considered parton species, since the corresponding local operators coincide with the off-diagonal elements of the energy-momentum tensor. In the forward limit one has simple relations to the momentum fraction and the total angular momentum carried by a specific parton species (as pointed out by Ji \cite{Ji:1996ek}). The first moment of the parity-odd quark GPD has been linked to the quark spin-orbit correlation recently (see Ref.~\cite{Lorce:2014mxa}). We use the framework of two-flavor covariant baryon chiral perturbation theory (BChPT)~\cite{Gasser:1987rb,Becher:1999he} to obtain analytic formulas describing the pion mass dependence of the first moments at full one-loop order ($\simeq\mathcal{O}(p^3)$, where $p$ represents both the scale of the pion mass and the momentum transfer, which are considered to be small in the low-energy regime). There already exist various heavy baryon calculations on the topic (e.g.~\cite{Arndt:2001ye,Belitsky:2002jp,Diehl:2006ya,Diehl:2006js,Ando:2006sk,Wang:2010hp,Detmold:2005pt}) and covariant leading one-loop calculations \cite{Dorati:2007bk,Moiseeva:2012zi}, where third-order effects are only partially included. This work is an extension of Ref.~\cite{Dorati:2007bk}. Advancing to third chiral order, the number of possible low-energy structures rises significantly. Eliminations via the equations of motion (EOM) will therefore be an important topic in Sec.~\ref{sect_operator_construction}. It turns out that the full one-loop result contains numerous new low-energy constants (LECs). In Sec.~\ref{sect_calculation} we therefore discuss two possibilities to reduce the number of free parameters if the available lattice data is not sufficient to pin down all LECs. The reduction of the results to the heavy BChPT version is presented in Sec.~\ref{sect_heavy_baryon_reduction}. We begin this article with a review of properties of chiral-even GPDs to establish a basis for the later analysis.

%
%

\section{Chiral-even GPDs \label{sect_chiral_even_GPDs}}

There are in total four different kinds of chiral-even GPDs: two quark distributions $F^q$, $\tilde{F}^q$, which are defined for each quark flavor separately, and two gluon distributions $F^g$, $\tilde{F}^g$, where $\tilde{F}^q$ and $\tilde{F}^g$ are (parton) spin dependent. The $n$th $x$ moments of these quantities are defined as the integral over $x$ from $-1$ to $1$, with the weight factor $x^n$ for quark and $x^{n-1}$ for gluon distributions. Mirroring the fact that gluons are their own antiparticles, $F^g$ is an even and $\tilde{F}^g$ an odd function of $x$ and all odd moments of $\tilde{F}^g$ vanish accordingly. The three GPDs with nonvanishing first moments are defined as \cite{Diehl:2003ny}
\begin{subequations}
 \begin{align}
\begin{split}
F^q &=\frac{1}{2} \int \limits_{-\infty}^\infty \frac{dz^-}{2 \pi} e^{i x \bar{p}^+ z^-} \langle p^\prime | \bar{q}(-z/2) \gamma^+ q(z/2) | p \rangle \\
 &= \frac{1}{2 \bar{p}^+} \bar{u}(p^\prime) \bigg[ H^q(x,\xi,t) \gamma^+  +E^q(x,\xi,t) \frac{i \sigma^{+ \alpha} \Delta_\alpha}{2 m_N} \bigg] u(p) \ ,
\end{split} \\ 
\begin{split}
 \tilde{F}^q &=  \frac{1}{2} \int \limits_{-\infty}^\infty \frac{dz^-}{2 \pi} e^{i x \bar{p}^+ z^-} \langle p^\prime | \bar{q}(-z/2) \gamma^+ \gamma_5 q(z/2) | p \rangle \\
 &= \frac{1}{2 \bar{p}^+} \bar{u}(p^\prime) \bigg[ \tilde{H}^q(x,\xi,t)   \gamma^+ \gamma_5  + \tilde{E}^q(x,\xi,t) \frac{\gamma_5 \Delta^+}{2 m_N} \bigg] u(p) \ ,
\end{split} \\ 
\begin{split}
  F^g &= \frac{1}{\bar{p}^+} \int \limits_{-\infty}^\infty \frac{dz^-}{2 \pi} e^{i x \bar{p}^+ z^-} \langle p^\prime | G^{+\mu}(-z/2) G_\mu^{\ +}(z/2) | p \rangle \\
 &= \frac{1}{2 \bar{p}^+} \bar{u}(p^\prime) \bigg[ H^g(x,\xi,t)  \gamma^+  +  E^g(x,\xi,t) \frac{i \sigma^{+ \alpha} \Delta_\alpha}{2 m_N} \bigg] u(p) \ ,
\end{split}
\end{align}
\end{subequations}
where the spin dependence of the nucleon states and spinors is not written out explicitly and $z$ is a lightlike vector with vanishing plus component (for the definition of plus and minus components see Ref.~\cite{Diehl:2003ny}; compare also~\cite{Ji:1997gm}). To assure gauge invariance the fields in the nonlocal operators are connected by Wilson lines which we do not write out explicitly. The kinematic variables are defined as
\begin{align}
  \bar{p} &= \frac{1}{2}(p^\prime+p) \ , & \Delta&=p^\prime-p \ ,  \nonumber \\
 \xi &= -\frac{\Delta^+}{2 \ \bar{p}^+} \ , & t&=\Delta^2 \ .
\end{align}
The first moments of the GPDs are related to the matrix elements of the local twist-two operators,
\begin{subequations} \label{def_local_operators}
\begin{align}
\mathcal{O}_{\mu\nu}^q &= \frac{1}{2} \, \mathbf{S} \, \bar{q} \gamma_\mu (i D_{\nu}^-) q \ , \\
\tilde{\mathcal{O}}_{\mu\nu}^q &= \frac{1}{2} \, \mathbf{S} \, \bar{q} \gamma_\mu \gamma_5 (i D_{\nu}^-)q \ , \\
\mathcal{O}_{\mu\nu}^g &= \mathbf{S} \, G_{\mu \alpha} G^\alpha_{\ \nu} \ ,
\end{align}
\end{subequations}
where $D_\mu^- = \Dr_\mu - \Dl_\mu$ and the operator $\mathbf{S}$ projects onto leading twist by symmetrizing in Lorentz indices and subtracting traces. When acting on an object with two open indices $\mathbf{S}$ has the explicit form
\begin{align}
 \mathbf{S} \, \mathcal{O}^{\mu\nu} = \mathbf{S}^{\mu \nu}_{\alpha \beta} \, \mathcal{O}^{\alpha\beta} = \frac{1}{2} \bigl( g^\mu_\alpha g^\nu_\beta +  g^\mu_\beta g^\nu_\alpha - \frac{2}{d} g^{\mu \nu} g_{\alpha \beta} \bigr) \mathcal{O}^{\alpha\beta} \ .
\end{align}
The nucleon-to-nucleon matrix elements of these local currents are accessible by lattice simulations and decompose into \cite{Diehl:2003ny}
\begin{subequations} \label{formfactordecomposition}
\begin{align}
\begin{split}
 \langle p^\prime | \mathcal{O}_{\mu\nu}^q  | p \rangle & = \mathbf{S} \, \bar{u}(p^\prime)  \biggl[
 \begin{aligned}[t] & \gamma_\mu \bar{p}_{\nu} A^q_{2,0}(t)  + \frac{i \sigma_{\mu \alpha} \Delta^\alpha}{2 m_N} \bar{p}_{\nu} B^q_{2,0}(t)\\
 &+\frac{\Delta_\mu \Delta_\nu}{m_N} C^q_{2}(t) \biggr] u(p) \ , 
\end{aligned}	 
\end{split} \raisetag{\baselineskip}  \\
\begin{split}
 \langle p^\prime | \tilde{\mathcal{O}}_{\mu\nu}^q  | p \rangle & =  \mathbf{S} \, \bar{u}(p^\prime)  \biggl[
 \begin{aligned}[t] &  \gamma_\mu \gamma_5  \bar{p}_{\nu} \tilde{A}^q_{2,0}(t)\\
  &+ \frac{\gamma_5 \Delta_\mu}{2 m_N} \bar{p}_\nu  \tilde{B}^q_{2,0}(t) \biggr] u(p) \ ,
\end{aligned}
\end{split}   \\
\begin{split}
 \langle p^\prime | \mathcal{O}_{\mu\nu}^g  | p \rangle & = \mathbf{S} \, \bar{u}(p^\prime) \biggl[
\begin{aligned}[t] & \gamma_\mu \bar{p}_{\nu} A^g_{2,0}(t)  + \frac{i \sigma_{\mu \alpha} \Delta^\alpha}{2 m_N} \bar{p}_{\nu} B^g_{2,0}(t)\\
 &+\frac{\Delta_\mu \Delta_\nu}{m_N} C^g_{2}(t) \biggr] u(p) \ . \raisetag{\baselineskip} 
\end{aligned}
\end{split}
\end{align}
\end{subequations}
The occurring generalized form factors are related to the GPD moments via
\begin{subequations}
\begin{align}
 \int \limits_{-1}^1 \!\!\! dx \, x \, H^q(x,\xi,t) & =  A^q_{2,0}(t) + 4 \xi^2 C^q_{2}(t)  \ , \\
 \int \limits_{-1}^1 \!\!\! dx \, x \, E^q(x,\xi,t) & =  B^q_{2,0}(t) - 4 \xi^2 C^q_{2}(t)  \ ,  \\
 \int \limits_{-1}^1 \!\!\! dx \, x \, \tilde{H}^q(x,\xi,t) & = \tilde{A}^q_{2,0}(t) \ , \\
 \int \limits_{-1}^1 \!\!\! dx \, x \, \tilde{E}^q(x,\xi,t) & = \tilde{B}^q_{2,0}(t) \ ,  \\
 \int \limits_{0}^1 \!\!\! dx \, H^g(x,\xi,t) & = A^g_{2,0}(t) + 4 \xi^2 C^g_{2}(t)  \ ,\\
 \int \limits_{0}^1 \!\!\! dx \, E^g(x,\xi,t) & = B^g_{2,0}(t) - 4 \xi^2 C^g_{2}(t) \ .
\end{align}
\end{subequations}
In the forward limit the form factor $A^p_{2,0}$ corresponds to the mean momentum fraction carried by a parton species $p$, while the sum $(A^p_{2,0} + B^p_{2,0})/2$ yields its total angular momentum in a longitudinally polarized nucleon (see e.g.~\cite{Ji:1996ek}). These relations are direct consequences of the equality between $\mathcal{O}^q_{\mu\nu}$/$\mathcal{O}^g_{\mu\nu}$ and the quark/gluon part of the Belinfante improved energy-momentum tensor. From the fact that all parton species together have to carry the (total angular) momentum of the nucleon one directly deduces sum rules for the generalized form factors:
\begin{subequations} \label{sum_rules}
 \begin{align}
 A^g_{2,0}(0)+\sum_q A^q_{2,0}(0)=1 \ , \\
 B^g_{2,0}(0)+\sum_q B^q_{2,0}(0)=0 \ .
\end{align}
\end{subequations}

%
%

\section{Operator construction \label{sect_operator_construction}}

In this section we construct the low-energy version of the local operators given in Eq.~\eqref{def_local_operators}. We do this by taking into account all possible combinations of chiral building blocks that have the desired properties under chiral rotations ($\hat{\chi}$), parity transformation ($\hat{p}$), charge ($\hat{c}$) and Hermitian ($\dagger$) conjugation. When considering the parity transformation of objects with open Lorentz indices we always give the transformation of the zero components.

\subsection{Symmetry properties of the chiral fields}

We will use one of the standard formulations of two-flavor BChPT (for general BChPT reviews, see e.g.~\cite{Bernard:1995dp,Bernard:2007zu}). However, we neglect all external fields from the beginning, since they are not of interest for our considerations. The pion fields are contained in
\begin{align}
 u=\exp \biggl( \frac{i}{2 F_0} \tau^a \pi^a \biggr) \ ,
\end{align}
where $\tau^1$, $\tau^2$ and $\tau^3$ are Pauli matrices and $F_0$ is the pion decay constant in the chiral limit, so that $F_\pi = F_0 + \mathcal{O}(m_\pi^2) \approx \unit{92}{\mega\electronvolt}$. The isospinor $\Psi$ contains the proton and the neutron fields. Let us from now on use $X\in\{R,L\}$, $\bar{L}=R$ and $\bar{R}=L$. Where they are not used as an index, $L$ and $R$ are meant to be elements of $\operatorname{SU}(2)_{L/R}$. Defining $u_R = u$ and $u_L = u^\dagger$ the transformation properties of $u$ are
\begin{subequations}
\begin{align}
 u_{X,fg} &\overset{\hat{\chi}}{\longrightarrow} \bigl( X u_{X} K^\dagger \bigr)_{fg} =\bigl( K u_{X} \bar{X}^\dagger \bigr)_{fg} \ , \\
 u_{X,fg} &\overset{\hat{p}}{\longrightarrow}  u_{\bar{X},fg} \ , \\
 u_{X,fg} &\overset{\hat{c}}{\longrightarrow}  u_{X,gf} \ , \\
 u_{X,fg} &\overset{\dagger}{\longrightarrow}  u_{\bar{X},gf} \ ,
\end{align}
\end{subequations}
with flavor indices $f,g$ and the so-called compensator field $K$, which is a common, nonlinear realization of chiral symmetry \cite{Coleman:1969sm,Gasser:1987rb}. For the nucleon field we have
\begin{subequations}
\begin{align}
 \Psi^\beta_{g} &\overset{\hat{\chi}}{\longrightarrow} \bigl( K \Psi^\beta \bigr)_g \ , &  \bar{\Psi}^\alpha_{f} &\overset{\hat{\chi}}{\longrightarrow} \bigl( \bar{\Psi}^\alpha K^\dagger \bigr)_f \ , \\
 \Psi^\beta_{g} &\overset{\hat{p}}{\longrightarrow} \bigl( \gamma_0 \Psi_{g} \bigr)^\beta \ , &  \bar{\Psi}^\alpha_{f} &\overset{\hat{p}}{\longrightarrow} \bigl( \bar{\Psi}_{f} \gamma_0 \bigr)^\alpha \ , \\
 \Psi^\beta_{g} &\overset{\hat{c}}{\longrightarrow} \bigl( \bar{\Psi}_{g} C \bigr)^\beta \ , &  \bar{\Psi}^\alpha_{f} &\overset{\hat{c}}{\longrightarrow} \bigl( C \Psi_{f} \bigr)^\alpha \ ,  \\
 \Psi^\beta_{g} &\overset{\dagger}{\longrightarrow} \bigl( \bar{\Psi}_{g} \gamma_0 \bigr)^\beta \ , &  \bar{\Psi}^\alpha_{f} &\overset{\dagger}{\longrightarrow} \bigl( \gamma_0 \Psi_{f} \bigr)^\alpha \ ,
\end{align}
\end{subequations}
where $C=i\gamma^2 \gamma^0$ is the charge conjugation matrix. The covariant derivative acting on a nucleon field is defined as
\begin{align}
 \Dr_\mu \Psi  &= (\partial_\mu + \Gamma_\mu) \Psi \ , & \bar{\Psi} \Dl_\mu &= \bar{\Psi} ( \overset{\leftarrow}{\partial}_\mu - \Gamma_\mu) \ ,
\end{align}
where $\Gamma_\mu$ is called the chiral connection and is given by
\begin{align}
 \Gamma_\mu &= \frac{1}{2} \bigl( u^\dagger \partial_\mu u + u \partial_\mu u^\dagger \bigr) \ .
\end{align}
Working with nucleon bilinears it is convenient to define left-right (covariant) derivatives that only act on the nucleon fields,
\begin{align}
  \bar{\Psi} A \Gamma D^\pm_\mu \Psi &= \bar{\Psi} \bigl(A \Gamma \Dr_\mu \pm \Dl_\mu A \Gamma  \bigr) \Psi \ ,
\end{align}
where $A$ and $\Gamma$ are placeholders for arbitrary chiral building blocks and Dirac structures. It is easy to show that the covariant derivative sandwiched between two nucleon spinors has the following transformation properties,
\begin{subequations}
\begin{align}
D_\mu^\pm &\overset{\hat{\chi}}{\longrightarrow} K D_\mu^\pm K^\dagger \ , \\
D_\mu^\pm &\overset{\hat{p}}{\longrightarrow} \eta^P_{D^\pm} D_\mu^\pm  \ , \\
D_\mu^\pm &\overset{\hat{c}}{\longrightarrow} \eta^{C}_{D^\pm} D_\mu^\pm \ , \\
D_\mu^\pm &\overset{\dagger}{\longrightarrow} \eta^{h}_{D^\pm} D_\mu^\pm \ ,
\end{align}
\end{subequations}
where $\eta^P_{D^\pm} \equiv 1$, $\eta^{C}_{D^\pm} \equiv \pm1$ and $\eta^{h}_{D^\pm} \equiv \pm1$ (the minus sign from interchanging the nucleon fields is not included in $\eta^C$). Note that $D^+_\mu$ has to be counted as small (first order) since the nucleon mass term drops out in the momentum difference. In a case where we have more than one covariant derivative we define a string of plus and minus derivatives as
\begin{align}
\begin{split} \label{def_multiple_derivatives}
 D_{\mu_1}^{\pm_1} \cdots D_{\mu_n}^{\pm_n} &= \sum_{k=0}^n \sum_{\sigma \in S_n} \frac{(\pm_{_{\sigma(1)}}1) \cdots (\pm_{_{\sigma(k)}}1)}{k!(n-k)!} \\ &\quad \times \Dl_{\mu_{\sigma(1)}} \cdots \Dl_{\mu_{\sigma(k)}} \Dr_{\mu_{\sigma(k+1)}} \cdots \Dr_{\mu_{\sigma(n)}} \ ,
\end{split}
\end{align}
where $S_n$ is the symmetric group of degree $n$. The symmetrization within the covariant derivatives acting to the left/right is possible due to the curvature relation given below (see Eq.~\eqref{curvature_relation}). The derivatives are meant to act on the nucleon fields only. We find the transformation properties,
\begin{subequations}
\begin{align}
D_{\mu_1}^{\pm_1} \cdots D_{\mu_n}^{\pm_n} &\overset{\hat{\chi}}{\longrightarrow} K D_{\mu_1}^{\pm_1} \cdots D_{\mu_n}^{\pm_n} K^\dagger \ , \\
D_{\mu_1}^{\pm_1} \cdots D_{\mu_n}^{\pm_n} &\overset{\hat{p}}{\longrightarrow} \eta^P_{D^{\pm_1}} \cdots \eta^P_{D^{\pm_n}} D_{\mu_1}^{\pm_1} \cdots D_{\mu_n}^{\pm_n}  \ , \\
D_{\mu_1}^{\pm_1} \cdots D_{\mu_n}^{\pm_n} &\overset{\hat{c}}{\longrightarrow} \eta^C_{D^{\pm_1}} \cdots \eta^C_{D^{\pm_n}} D_{\mu_1}^{\pm_1} \cdots D_{\mu_n}^{\pm_n} \ , \\
D_{\mu_1}^{\pm_1} \cdots D_{\mu_n}^{\pm_n} &\overset{\dagger}{\longrightarrow} \eta^h_{D^{\pm_1}} \cdots \eta^h_{D^{\pm_n}} D_{\mu_1}^{\pm_1} \cdots D_{\mu_n}^{\pm_n} \ .
\end{align}
\end{subequations}
For the case of two derivatives Eq.~\eqref{def_multiple_derivatives} reads
\begin{align}
\begin{split}
   &\bar{\Psi} A \Gamma D^{\pm_1}_\mu D^{\pm_2}_\nu \Psi \\ 
   &\,=\bar{\Psi} \Bigl(A \Gamma \frac{1}{2}(\Dr_\mu \Dr_\nu+\Dr_\nu \Dr_\mu) \pm_1 \Dl_\mu A \Gamma \Dr_\nu \\
   &\,\quad\pm_2 \Dl_\nu A \Gamma \Dr_\mu + (\pm_1 1)(\pm_2 1) \frac{1}{2}(\Dl_\mu \Dl_\nu+\Dl_\nu \Dl_\mu)  A \Gamma  \Bigr) \Psi \ .
\end{split}
\end{align}
The chiral vielbein $u_\mu$ and the quark mass insertions $\chi^\pm$ are defined as
\begin{align}
  u_\mu &= i \bigl( u^\dagger \partial_\mu u - u \partial_\mu u^\dagger \bigr) \ , \\
  \chi^\pm &= \bigl( u^\dagger \chi u^\dagger \pm u \chi^\dagger u \bigr) \ ,
\end{align}
where $\chi=2 B_0 \mathcal{M}$ includes the quark mass matrix. For $A \in \{u_\mu,\chi^\pm\}$ we have the transformation properties
\begin{subequations}
\begin{align}
A_{fg} &\overset{\hat{\chi}}{\longrightarrow} \bigl( K A K^\dagger \bigr)_{fg} \ , \\
A_{fg} &\overset{\hat{p}}{\longrightarrow} \eta^P_A A_{fg}  \ , \\
A_{fg}  &\overset{\hat{c}}{\longrightarrow} \eta^C_A A_{gf} \ , \\
A_{fg}  &\overset{\dagger}{\longrightarrow} \eta^h_A A_{gf} \ ,
\end{align}
\end{subequations}
where the corresponding $\eta$'s can be found in Table~\ref{symmetry_properties}. In the same table the reader finds the transformation properties formally assigned to the covariant derivative acting on those building blocks $D_\mu A = \partial_\mu A + [\Gamma_\mu,A]$. For strings of building blocks we obtain
\begin{subequations}
\begin{align}
\eta^P_{\{A,B\}_\pm} &= \eta^P_A \eta^P_B \ , \\
\eta^C_{\{A,B\}_\pm} &= \pm \eta^C_A \eta^C_B \ , \\
\eta^h_{\{A,B\}_\pm} &= \pm \eta^h_A \eta^h_B \ .
\end{align}
\end{subequations}
Finally we define for the elements of the Clifford algebra,
\begin{subequations}
\begin{align}
\Gamma &= \eta^P_\Gamma \gamma_0 \Gamma \gamma_0 \ , \\
\Gamma^T &= \eta^C_\Gamma C \Gamma C \ , \\
\Gamma^\dagger &= \eta^h_\Gamma \gamma_0 \Gamma \gamma_0 \ .
\end{align}
\end{subequations}
\begin{table}[tp]
\caption{\label{symmetry_properties} The constants $\eta^P$, $\eta^C$ and $\eta^h$ characterizing the symmetry properties of the basic operators, elements of the Clifford algebra and the Levi-Civita symbol.}
\begin{ruledtabular}
\begin{tabular}{l r r r}
$A$ & $\eta^P_A$ & $\eta^C_A$ & $\eta^h_A$ \\ \hline
$1$ & $1$ & $1$ & $1$ \\
$u_\mu$ & $-1$ & $ 1$ & $ 1$ \\
$\chi^\pm$ & $\pm1$ & $ 1$ & $ \pm1$ \\
$D_\mu^\pm$ & $1$ & $\pm 1$ & $\pm 1$ \\
$D_\mu \text{ (acting on pion field)}$ & $1$ & $ 1 $ & $ 1$ \\
$\mathds{1}$ & $1$ & $-1$ & $1$ \\
$\gamma_5$ & $-1$ & $-1$ & $-1$ \\
$\gamma_\mu$ & $1$ & $1$ & $1$ \\
$\gamma_\mu \gamma_5$ & $-1$ & $-1$ & $1$ \\
$\sigma_{\mu\nu}$ & $1$ & $1$ & $1$ \\
$\epsilon_{\mu\nu\rho\sigma}$ & $-1$ & $1$ & $1$
\end{tabular}
\end{ruledtabular}
\end{table}%
\subsection{Symmetry properties of the operators}

For the following construction it is convenient to define operators with arbitrary quark flavors by
\begin{subequations}
\begin{align}
\mathcal{O}_{\mu\nu}^Q &= \frac{1}{2} \, \mathbf{S} \, \bar{q}_f \gamma_\mu (i D_\nu^-) Q_{fg} q_g \ , \\
\tilde{\mathcal{O}}_{\mu \nu}^Q &= \frac{1}{2} \, \mathbf{S} \, \bar{q}_f \gamma_\mu \gamma_5 (i D_\nu^-) Q_{fg} q_g \ ,
\end{align}
\end{subequations}
where $Q$ is a $2\times2$ matrix and $f,g$ are flavor indices. Being chiral even, the operators can be split in parts that contain left- or right-handed quarks ($q_X = \gamma_X q_X$, $\gamma_{R/L}=(\mathds{1}\pm\gamma_5)/2$) exclusively,
\begin{subequations}
\begin{align}
 \mathcal{O}_{\mu\nu}^Q &= Q_{fg} \bigl( \mathcal{O}_{R, \mu \nu  , fg} + \mathcal{O}_{L, \mu \nu  , fg}  \bigr)  \ , \\
 \tilde{\mathcal{O}}_{\mu\nu}^Q &= Q_{fg} \bigl( \mathcal{O}_{R, \mu \nu  , fg} - \mathcal{O}_{L, \mu \nu  , fg}  \bigr)  \ ,
\end{align}
\end{subequations}
where the operators $\mathcal{O}_X$, $X \in \{R,L\}$ are given by
\begin{align}
 \mathcal{O}_{X,\mu \nu , fg} &= \frac{1}{2} \mathbf{S} \bar{q}_{X f} \gamma_\mu (i D_\nu^-) q_{X g} \ .
\end{align}
The transformation properties of the (right- and left-handed) quark and antiquark fields under chiral rotations, parity transformations, charge and Hermitian conjugation,
\begin{subequations}  \label{tp_quarkfields}
\begin{align}
 q^\beta_{Xg} &\overset{\hat{\chi}}{\longrightarrow} \bigl( X q^\beta_{X} \bigr)_g \ , &  \bar{q}^\alpha_{Xf} &\overset{\hat{\chi}}{\longrightarrow} \bigl( \bar{q}^\alpha_{X} X^\dagger \bigr)_f \ , \\
 q^\beta_{Xg} &\overset{\hat{p}}{\longrightarrow} \bigl( \gamma_0 q_{\bar{X}g} \bigr)^\beta \ , &  \bar{q}^\alpha_{Xf} &\overset{\hat{p}}{\longrightarrow} \bigl( \bar{q}_{\bar{X}f} \gamma_0 \bigr)^\alpha \ , \\
 q^\beta_{Xg} &\overset{\hat{c}}{\longrightarrow} \bigl( \bar{q}_{\bar{X}g} C \bigr)^\beta \ , &  \bar{q}^\alpha_{Xf} &\overset{\hat{c}}{\longrightarrow} \bigl( C q_{\bar{X}f} \bigr)^\alpha \ ,  \\
 q^\beta_{Xg} &\overset{\dagger}{\longrightarrow} \bigl( \bar{q}_{Xg} \gamma_0 \bigr)^\beta \ , &  \bar{q}^\alpha_{Xf} &\overset{\dagger}{\longrightarrow} \bigl( \gamma_0 q_{Xf} \bigr)^\alpha \ ,
\end{align}
\end{subequations}
yield the following symmetry properties for the composite operators:
\begin{subequations} \label{symmetry_properties_operator}
\begin{align} 
\mathcal{O}_{X, \mu \nu, fg} &\overset{\hat{\chi}}{\longrightarrow} X_{g g^\prime} \mathcal{O}_{X, \mu \nu, f^\prime g^\prime} X^\dagger_{f^\prime f} \ , \\
\mathcal{O}_{X, \mu \nu, fg} &\overset{\hat{p}}{\longrightarrow} \mathcal{O}_{\bar{X}, \mu \nu, fg} \ , \\
\mathcal{O}_{X, \mu \nu, fg} &\overset{\hat{c}}{\longrightarrow} \mathcal{O}_{\bar{X}, \mu \nu, gf} \ , \\
\mathcal{O}_{X, \mu \nu, fg} &\overset{\dagger}{\longrightarrow} \mathcal{O}_{X, \mu \nu, gf} \ .
\end{align}
\end{subequations}
These transformation properties have to be reproduced by the low-energy version of the currents and will guide the construction performed in the following sections.

\subsection{Mesonic sector}

In the mesonic sector we can easily write down the operator in terms of chiral fields which have the correct properties under chiral rotations,
\begin{align}
\begin{split}
\mathcal{O}_{X, \mu \nu, fg} &= \mathbf{S} \; \sum_{A} L_{X,A} \bigl( u_X A_{\mu \nu} u_{\bar{X}} \bigr)_{gf}\\
&\quad+  \mathds{1}_{gf}\;   \mathbf{S} \;   \sum_{A} L^s_{X,A} \tr{ u_X A_{\mu \nu} u_{\bar{X}} } \ ,
\end{split}
\end{align}
where the $A$'s have to be combinations of chiral building blocks. To comply with the remaining symmetry properties given in Eq.~\eqref{symmetry_properties_operator} we have to demand that the LECs fulfill the following constraints:
\begin{align}
L_{\bar{X},A} &= L_{X,A} \eta^P_A  \in
\begin{cases}
\{0\} &  \text{if } \eta^C_A \eta^P_A  = - 1 \\
\mathds{R} &  \text{if } \eta^h_A = 1 \\
\mathds{I} &  \text{if } \eta^h_A = -1 
\end{cases} \ .
\end{align}
Up to second (chiral) order there is only one possible structure for $A$, which is $u_{ \mu} u_{\nu}$. Defining the LECs such that isosinglet and -triplet are neatly separated ($l=L_{R,A}$ and $l^s=L^s_{R,A}+L_{R,A}/2$), we find for the mesonic part of the complete operators,
\begin{subequations}
\begin{align}
\begin{split}
  \mathcal{O}_{\mu\nu}^Q \Bigr|_{\pi,2} &= \mathbf{S} \tr{\bigl(l \tilde{Q} + l^s \tr{Q} \bigr) \bigl( u u_{ \mu} u_{\nu } u^\dagger +  u^\dagger u_{ \mu} u_{\nu } u \bigr) } \ .
\end{split} \\
\begin{split}
  \tilde{\mathcal{O}}_{\mu\nu}^Q\Bigr|_{\pi,2} &= \mathbf{S} \tr{ l \tilde{Q}  \bigl( u u_{ \mu} u_{\nu } u^\dagger -  u^\dagger u_{ \mu} u_{\nu } u \bigr) } \ ,
\end{split}
\end{align}
\end{subequations}
where
\begin{align}
 \tilde{Q}\equiv Q - \frac{1}{2} \tr{Q} \ .
\end{align}
These structures are of second chiral order. We will not discuss higher-order structures since they do not contribute to our calculation.

\subsection{Nucleon sector}

A power counting analysis of all possible Feynman diagrams (compare Fig.~\ref{FD_nGPDs}) yields that only zeroth- and first-order $N\pi\pi N$ and $N \pi N$ operator insertions contribute to graphs \subref{fd:subfigb}, \subref{fd:subfigd} and \subref{fd:subfige} at full one-loop level. Thus, in the nucleon sector, we only have to construct second- and third-order operator structures without additional pions. Working in the limit of exact isospin symmetry, this leaves us, aside from covariant derivatives acting on the nucleon fields, with only three possible chiral building blocks $\mathds{1}$, $\tr{\chi^+}\mathds{1}$ and $u^\mu$, where the chiral vielbein only occurs at most once in operators of first chiral order. As a first step we write down all structures with correct behavior under chiral transformations,
\begin{align}
\begin{split}
 \mathcal{O}_{X, \mu \nu, fg} &= \smashoperator{ \sum_{(\mathbf{A})_{\mu \nu},\pm}} L^{\pm}_{X,\mathbf{A}} \biggl(
\begin{aligned}[t] &
  \bigl( \bar{\Psi} A_1 u_{\bar{X}} \bigr)_f A_\Gamma \bigl( u_X A_2 \Psi \bigr)_g\\
& \pm \bigl( \bar{\Psi} A_2 u_{\bar{X}} \bigr)_f A_\Gamma \bigl( u_X A_1 \Psi \bigr)_g  \biggr)
\end{aligned} \\
&\quad+ \smashoperator{ \sum_{(\mathbf{A})_{\mu \nu}}} L^{s}_{X,\mathbf{A}} \bigl( u_X A_1 u_{\bar{X}} \bigr)_{gf} \bigl( \bar{\Psi} A_2 A_\Gamma \Psi \bigr) \ ,
\end{split}
\end{align}
with $\mathbf{A} \equiv A_1 \otimes A_2 \otimes A_\Gamma$, where $A_1$ and $A_2$ are chiral building blocks, while $A_\Gamma$ contains elements of the Clifford algebra and derivatives acting on the nucleon fields. In order to have correct properties under parity transformation, charge and Hermitian conjugation the LECs have to fulfill the following relations:
\begin{subequations}
\begin{align}
\begin{split}
  L^{\pm}_{\bar{X},\mathbf{A}} &= L^{\pm}_{X,\mathbf{A}} \eta^P_{A_1} \eta^P_{A_2} \eta^P_{A_\Gamma} \\
 &\in \begin{cases} \{0\} &  \text{if } \eta^C_{A_1} \eta^C_{A_2} \eta^C_{A_\Gamma} \eta^P_{A_1} \eta^P_{A_2} \eta^P_{A_\Gamma}= \pm 1 \\ \mathds{R} &  \text{if } \eta^h_{A_1} \eta^h_{A_2} \eta^h_{A_\Gamma}= \pm 1 \\ \mathds{I} &  \text{if } \eta^h_{A_1} \eta^h_{A_2} \eta^h_{A_\Gamma}  = \mp 1 \end{cases} \ ,
\end{split} \\
\begin{split}
  L^{s}_{\bar{X},\mathbf{A}} &= L^{s}_{X,\mathbf{A}} \eta^P_{A_1} \eta^P_{A_2} \eta^P_{A_\Gamma} \\
 &\in \begin{cases} \{0\} &  \text{if } \eta^C_{A_1} \eta^C_{A_2} \eta^C_{A_\Gamma} \eta^P_{A_1} \eta^P_{A_2} \eta^P_{A_\Gamma}= + 1 \\ \mathds{R} &  \text{if } \eta^h_{A_1} \eta^h_{A_2} \eta^h_{A_\Gamma}= + 1 \\ \mathds{I} &  \text{if } \eta^h_{A_1} \eta^h_{A_2} \eta^h_{A_\Gamma}  = - 1 \end{cases} \ .
\end{split}
\end{align}
\end{subequations}
For the actual construction this form of the operator is quite unhandy. It is more convenient to treat the parts of the operator containing different chiral building blocks separately,
\begin{table}[tp]
\caption{\label{Dstructures} Elements of $\mathcal{D}_\mu^{(0)}$, $\mathcal{D}^{(0)}_{\{ \mu \nu \}}$, $\mathcal{D}^{(1)}_{\{ \mu \nu \}}$, $\mathcal{D}^{(2)}_{\{ \mu \nu \}}$, $\mathcal{D}^{(3)}_{\{ \mu \nu \}}$ and $\mathcal{D}^{(0)}_{\{ \mu \nu \} \alpha }$ and their symmetry properties. All operators in the second column still have to by symmetrized in $\mu$ and $\nu$ by $\mathbf{S}$. }
\begin{ruledtabular}
\begin{tabular}{l l r r r}
& $A$  & $\eta^P_A$ & $\eta^C_A$ & $\eta^h_A$  \\ \hline
$A \in \mathcal{D}^{(0)}_\mu$
			      &$\gamma_5 D_\mu^-$ & $-1$ & $+1$ & $+1$ \\
			      &$\gamma_\mu$ & $+1$ & $+1$ & $+1$ \\
			      &$\gamma_\mu \gamma_5$ & $-1$ & $-1$ & $+1$ \\ \hline
$A \in \mathcal{D}^{(0)}_{\{ \mu \nu \}}$
			      &$\gamma_5 D_\mu^- D_\nu^-$ & $-1$ & $-1$ & $-1$ \\
                              &$\gamma_\mu D_\nu^-$ & $+1$ & $-1$ & $-1$ \\
                              &$\gamma_\mu \gamma_5 D_\nu^-$ & $-1$ & $+1$ & $-1$ \\ \hline
$A \in \mathcal{D}^{(1)}_{\{ \mu \nu \}}$
			      &$\gamma_5 D_\mu^+ D_\nu^-$ & $-1$ & $+1$ & $+1$ \\
                              &$\gamma_\mu D_\nu^+$ & $+1$ & $+1$ & $+1$ \\
                              &$\gamma_\mu \gamma_5 D_\nu^+$ & $-1$ & $-1$ & $+1$ \\
                              &$\sigma_{\mu \beta} D^{+ \beta} D_\nu^-$ & $+1$ & $-1$ & $-1$ \\ \hline
$A \in \mathcal{D}^{(2)}_{\{ \mu \nu \}}$
			      &$D_\mu^+ D_\nu^+$ & $+1$ & $-1$ & $+1$ \\
                              &$\gamma_5 D_\mu^+ D_\nu^+$ & $-1$ & $-1$ & $-1$ \\
                              &$\sigma_{\mu \beta} D^{+ \beta} D_\nu^+$ & $+1$ & $+1$ & $+1$ \\ \cline{2-5}
                              &$ + \mathcal{D}^{(0)}_{\{ \mu \nu \}} \ D^+ \! \! \cdot \! D^+$ &&&\\ \hline
$A \in \mathcal{D}^{(3)}_{\{ \mu \nu \}}$
			      &$ = \mathcal{D}^{(1)}_{\{ \mu \nu \}} \ D^+ \! \! \cdot \! D^+$ &&&\\ \hline
$A \in \mathcal{D}^{(0)}_{\{ \mu \nu \} \alpha }$
			      &$\gamma_5 D_\mu^- D_\nu^- D_\alpha^-$ & $-1$ & $+1$ & $+1$ \\
                              &$\gamma_\mu D_\nu^- D_\alpha^-$ & $+1$ & $+1$ & $+1$ \\
                              &$\gamma_\alpha D_\mu^- D_\nu^-$ & $+1$ & $+1$ & $+1$ \\                              
                              &$\gamma_\mu \gamma_5 D_\nu^- D_\alpha^-$ & $-1$ & $-1$ & $+1$ \\ 
                              &$\gamma_\alpha \gamma_5 D_\mu^- D_\nu^-$ & $-1$ & $-1$ & $+1$ \\
                              &$\sigma_{\mu \alpha} D_\nu^-$ & $+1$ & $-1$ & $-1$ \\    
                              &$\epsilon_{\mu\alpha\rho\sigma}\sigma^{\rho \sigma} D_\nu^-$ & $-1$ & $-1$ & $-1$ \\                              
\end{tabular}
\end{ruledtabular}
\end{table}%
\begin{subequations}
\begin{align}
\begin{split}
  \mathcal{O}_{X, \mu \nu, fg}^{\mathds{1},(n)} &= \smashoperator[l]{\sum \limits_{A_{\mu \nu}\in\mathcal{D}^{(n)}_{\{\mu \nu\}}}} \!\! \biggl(
\!\! \begin{aligned}[t]
& L^{\mathds{1}}_{X,A} \bigl( \bar{\Psi}  u_{\bar{X}} \bigr)_f A_{\mu\nu} \bigl( u_X \Psi \bigr)_g\\
&+  L^{\mathds{1},s}_{X,A} \, \mathds{1}_{gf} \, \bigl( \bar{\Psi} A_{\mu \nu} \Psi \bigr) \biggr)  \ , 
\end{aligned}
\end{split}    \\
\begin{split}
\mathcal{O}_{X, \mu \nu, fg}^{\chi^+,(n \ge 2)} &= \tr{\chi^+} \smashoperator[l]{\sum_{A_{\mu \nu}\in\mathcal{D}^{(n-2)}_{\{\mu \nu\}}}} \biggl(
\!\! \begin{aligned}[t]
&  L^{\chi^+}_{X,A}  \bigl( \bar{\Psi}  u_{\bar{X}} \bigr)_f A_{\mu\nu} \bigl( u_X \Psi \bigr)_g \\
&+ L^{\chi^+,s}_{X,A} \, \mathds{1}_{gf} \, \bigl( \bar{\Psi} A_{\mu \nu} \Psi \bigr) \biggr)  \ , 
\end{aligned}
\end{split} \raisetag{20pt}    \\
\begin{split}
 \mathcal{O}_{X, \mu \nu, fg}^{u,(1)} &= \smashoperator[l]{\sum_{ \substack{ A_{\mu \nu \alpha} \in \mathcal{D}^{(0)}_{\{\mu \nu\}\alpha} \\ A_{\mu \nu \alpha} \in g_{\alpha \{ \mu}^{\phantom{(0)}} \mathcal{D}^{(0)}_{\nu \}}}}} \!\!\!\! \biggl(
\!\! \begin{aligned}[t]
 & L^{u}_{X,A} \Bigl( 
  \begin{aligned}[t]
      &\bigl( \bar{\Psi} u^\alpha u_{\bar{X}} \bigr)_f A_{\mu\nu \alpha} \bigl( u_X \Psi \bigr)_g \\
      &\!\!\!\!\!\!+ \eta^C_A \eta^P_A \bigl( \bar{\Psi}  u_{\bar{X}} \bigr)_f A_{\mu\nu \alpha} \bigl( u_X u^\alpha \Psi \bigr)_g \Bigr)
 \end{aligned} \\
 &+ L^{u,s1}_{X,A} \, \mathds{1}_{gf} \, \bigl( \bar{\Psi} u^\alpha A_{\mu \nu \alpha} \Psi \bigr)\\
 &+ L^{u,s2}_{X,A} \, \bigl(u_X u^\alpha u_{\bar{X}}\bigr)_{gf} \bigl( \bar{\Psi} A_{\mu \nu \alpha} \Psi \bigr) \biggr) \ ,
\end{aligned}
 \end{split}
\end{align}
\end{subequations}
where the number in brackets denotes the chiral order. For the LECs we then have the following relations:
\begin{subequations}
\begin{align}
  L^{\mathds{1}(,s)}_{\bar{X},A} & = L^{\mathds{1}(,s)}_{X,A} \eta^P_{A} \in \begin{cases} \{0\} &  \text{if } \eta^C_A \eta^P_A= + 1 \\ \mathds{R} &  \text{if } \eta^h_A= + 1 \\ \mathds{I} &  \text{if } \eta^h_A  = - 1 \end{cases} \ ,    \\
  L^{\chi^+(,s)}_{\bar{X},A} & = L^{\chi^+(,s)}_{X,A} \eta^P_{A} \in \begin{cases} \{0\} &  \text{if } \eta^C_A \eta^P_A= + 1 \\ \mathds{R} &  \text{if } \eta^h_A= + 1 \\ \mathds{I} &  \text{if } \eta^h_A  = - 1 \end{cases} \ ,    \\
  L^{u}_{\bar{X},A} & = L^{u}_{X,A} \eta^P_{A} \in \begin{cases} \mathds{R} &  \text{if } \eta^h_A= + \eta^C_A \eta^P_A \\ \mathds{I} &  \text{if } \eta^h_A  = - \eta^C_A \eta^P_A \end{cases} \ , \\
  L^{u,s1/s2}_{\bar{X},A} & = L^{u,s1/s2}_{X,A} \eta^P_{A} \in \begin{cases} \{0\} &  \text{if } \eta^C_A \eta^P_A= - 1 \\ \mathds{R} &  \text{if } \eta^h_A= + 1 \\ \mathds{I} &  \text{if } \eta^h_A  = - 1 \end{cases} \ .   
\end{align}
\end{subequations}
The elements of the $\mathcal{D}$'s and their symmetry properties are collected in Table~\ref{Dstructures}. To reduce the number of possible terms it is essential to use the (free) EOM,
\begin{align} \label{EOM}
  i \overset{\rightarrow}{\slashed{D}} \Psi &\overset{\cdot}{=} m \Psi \ , & \bar{\Psi} i\overset{\leftarrow}{\slashed{D}} &\overset{\cdot}{=} - \bar{\Psi} m \ ,
\end{align}
where the dot over the equal sign means up to higher order. Using Eq.~\eqref{EOM} one finds two identities,
\begin{subequations}
\begin{align}
  \bar{\Psi} \Gamma \gamma^\beta \Dr_\beta \Psi &= \bar{\Psi} \biggl( \frac{1}{2} \{ \Gamma, \gamma^\beta\} +  \frac{1}{2} [ \Gamma, \gamma^\beta ]  \biggr) \Dr_\beta \Psi \overset{\cdot}{=} - i m \bar{\Psi} \Gamma \Psi \ ,\\
  \bar{\Psi} \gamma^\beta \Gamma  \Dl_\beta \Psi &= \bar{\Psi} \biggl( \frac{1}{2} \{ \Gamma, \gamma^\beta\} -  \frac{1}{2} [ \Gamma, \gamma^\beta ]  \biggr) \Dl_\beta \Psi \overset{\cdot}{=} i m \bar{\Psi} \Gamma \Psi \ ,
\end{align}
\end{subequations}
which can be rewritten as
\begin{subequations}
\begin{align}
 \bar{\Psi} \biggl( \frac{1}{2} \{ \Gamma, \gamma^\beta\} D^+_\beta +  \frac{1}{2} [ \Gamma, \gamma^\beta ] D^-_\beta  \biggr) \Psi &\overset{\cdot}{=} 0 \ , \\
 \bar{\Psi} \biggl( \frac{1}{2} \{ \Gamma, \gamma^\beta\} D^-_\beta +  \frac{1}{2} [ \Gamma, \gamma^\beta ] D^+_\beta  \biggr) \Psi &\overset{\cdot}{=} - 2 i m \bar{\Psi} \Gamma \Psi \ ,
\end{align}
\end{subequations}
\begin{table}[tp]
\caption{\label{LE_Operators} Structures contributing to the full one-loop calculation. We only write down terms that survive the EOM eliminations discussed in the text and the symmetrization by $\mathbf{S}$ later on.}
\begin{ruledtabular}
\begin{tabular}{l l l l}
$n$& $i$  &$\mathcal{O}^{n,i}_{\mu\nu}$ & $\tilde{\mathcal{O}}^{n,i}_{\mu\nu}$\\ \hline
$0$
			      &1 & $ u^{Q,+s}_{0,1} \gamma_\mu i D_\nu^-$  & $u^{Q,-}_{0,1}  \gamma_\mu i D_\nu^-$ \\
			      &2 & $ u^{Q,-}_{0,2} \gamma_\mu \gamma_5 i D_\nu^-$  & $ u^{Q,+s}_{0,2}  \gamma_\mu \gamma_5 i D_\nu^-$ \\ \hline
$1$
			      &1 &  $u^{Q,-}_{1,1} \gamma_5 D_\mu^+ D_\nu^-$  &  $ u^{Q,+s}_{1,1} \gamma_5 D_\mu^+ D_\nu^-$  \\
			      &2 &  $u^{Q,+s}_{1,2} \sigma_\mu^{\ \alpha} D_\alpha^+ i D_\nu^-$ & $u^{Q,-}_{1,2} \sigma_\mu^{\ \alpha} D_\alpha^+ i D_\nu^-$\\ \cline{2-4}
			      &3 &  $\bigl\{ u_\mu , u^{Q,+}_{1,3} \bigr\}_- \gamma_5  i D_\nu^-$ &  $\bigl\{ u_\mu , u^{Q,-}_{1,3} \bigr\}_-  \gamma_5  i D_\nu^- $\\
			      &4 &  $\bigl\{ u_\mu , u^{Q,-}_{1,4} \bigr\}_+ \gamma_\nu$ & $ \bigl\{ u_\mu , u^{Q,+s}_{1,4} \bigr\}_+  \gamma_\nu$ \\
			      &5 &  $\tr{u_\mu u^{Q,-}_{1,5}}  \gamma_\nu $ & $\tr{u_\mu u^{Q,+}_{1,5}} \gamma_\nu$ \\
			      &6 &  $\bigl\{ u_\mu , u^{Q,+s}_{1,6} \bigr\}_+  \gamma_\nu \gamma_5$& $ \bigl\{ u_\mu , u^{Q,-}_{1,6} \bigr\}_+ \gamma_\nu \gamma_5 $  \\
			      &7 &  $\tr{u_\mu u^{Q,+}_{1,7}} \gamma_\nu \gamma_5$& $\tr{u_\mu u^{Q,-}_{1,7}} \gamma_\nu \gamma_5$   \\ \cline{2-4}
			      &8 &  $ \bigl\{ u^\alpha , u^{Q,+}_{1,8} \bigr\}_- \gamma_5 i D_\mu^- D_\nu^- D_\alpha^-$& $\bigl\{ u^\alpha , u^{Q,-}_{1,8} \bigr\}_- \gamma_5 i D_\mu^- D_\nu^- D_\alpha^-$ \\
			      &9 &  $ \bigl\{ u^\alpha , u^{Q,-}_{1,9} \bigr\}_+ \gamma_\mu D_\nu^- D_\alpha^-$& $ \bigl\{ u^\alpha , u^{Q,+s}_{1,9} \bigr\}_+ \gamma_\mu D_\nu^- D_\alpha^-$  \\
			      &10&  $ \tr{ u^\alpha u^{Q,-}_{1,10}} \gamma_\mu D_\nu^- D_\alpha^-$& $ \tr{ u^\alpha u^{Q,+}_{1,10}} \gamma_\mu D_\nu^- D_\alpha^-$  \\
			      &11&  $ \bigl\{ u^\alpha , u^{Q,-}_{1,11} \bigr\}_+ \gamma_\alpha D_\mu^- D_\nu^-$&  $ \bigl\{ u^\alpha , u^{Q,+s}_{1,11}\bigr\}_+ \gamma_\alpha D_\mu^- D_\nu^-$ \\
			      &12&  $ \tr{ u^\alpha u^{Q,-}_{1,12}} \gamma_\alpha D_\mu^- D_\nu^-$&  $ \tr{ u^\alpha u^{Q,+}_{1,12}} \gamma_\alpha D_\mu^- D_\nu^-$ \\
			      &13&  $ \bigl\{ u^\alpha , u^{Q,+s}_{1,13} \bigr\}_+ \gamma_\mu \gamma_5 D_\nu^- D_\alpha^-$&  $ \bigl\{ u^\alpha , u^{Q,-}_{1,13} \bigr\}_+ \gamma_\mu \gamma_5 D_\nu^- D_\alpha^-$ \\
			      &14&  $ \tr{ u^\alpha u^{Q,+}_{1,14}} \gamma_\mu \gamma_5 D_\nu^- D_\alpha^-$& $ \tr{ u^\alpha u^{Q,-}_{1,14}} \gamma_\mu \gamma_5 D_\nu^- D_\alpha^-$ \\      
			      &15&  $ \bigl\{ u^\alpha , u^{Q,+s}_{1,15} \bigr\}_+ \gamma_\alpha \gamma_5 D_\mu^- D_\nu^-$& $ \bigl\{ u^\alpha , u^{Q,-}_{1,15} \bigr\}_+ \gamma_\alpha \gamma_5 D_\mu^- D_\nu^-$ \\
			      &16&  $ \tr{ u^\alpha u^{Q,+}_{1,16}} \gamma_\alpha \gamma_5 D_\mu^- D_\nu^-$& $ \tr{ u^\alpha u^{Q,-}_{1,16}} \gamma_\alpha \gamma_5 D_\mu^- D_\nu^-$  \\
			      &17&  $ \bigl\{ u^\alpha , u^{Q,-}_{1,17} \bigr\}_- \sigma_{\mu \alpha} D_\nu^-$&  $ \bigl\{ u^\alpha , u^{Q,+}_{1,17} \bigr\}_- \sigma_{\mu \alpha} D_\nu^-$ \\
			      &18&  $ \bigl\{ u^\alpha , u^{Q,+s}_{1,18} \bigr\}_+ \epsilon_{\mu\alpha\rho\sigma}\sigma^{\rho \sigma} i D_\nu^-$& $ \bigl\{ u^\alpha , u^{Q,-}_{1,18}\bigr\}_+ \epsilon_{\mu\alpha\rho\sigma}\sigma^{\rho \sigma} i D_\nu^-$ \\
			      &19&  $ \tr{ u^\alpha  u^{Q,+}_{1,19}} \epsilon_{\mu\alpha\rho\sigma}\sigma^{\rho \sigma} i D_\nu^-$& $ \tr{ u^\alpha  u^{Q,-}_{1,19}} \epsilon_{\mu\alpha\rho\sigma}\sigma^{\rho \sigma} i D_\nu^-$ \\ \hline
$2$
			      &1 &  $u^{Q,+s}_{2,1} D_\mu^+ D_\nu^+$   & $u^{Q,-}_{2,1} D_\mu^+ D_\nu^+$   \\
			      &2 &  $u^{Q,+s}_{2,2} \gamma_\mu i D_\nu^- \tr{\chi^+} $ &  $ u^{Q,-}_{2,2} \gamma_\mu i D_\nu^- \tr{\chi^+} $\\
			      &3 &  $u^{Q,+s}_{2,3} \gamma_\mu i D_\nu^- D^+ \! \! \cdot \! D^+$& $u^{Q,-}_{2,3} \gamma_\mu i D_\nu^- D^+ \! \! \cdot \! D^+$ \\
			      &4 &  $u^{Q,-}_{2,4} \gamma_\mu \gamma_5 i D_\nu^- \tr{\chi^+}$& $ u^{Q,+s}_{2,4}  \gamma_\mu \gamma_5 i D_\nu^- \tr{\chi^+}$  \\
			      &5 &  $u^{Q,-}_{2,5} \gamma_\mu \gamma_5 i D_\nu^- D^+ \! \! \cdot \! D^+$& $u^{Q,+s}_{2,5} \gamma_\mu \gamma_5 i D_\nu^- D^+ \! \! \cdot \! D^+$ \\ \hline
$3$
			      &1 & $u^{Q,-}_{3,1} \gamma_5 D_\mu^+ D_\nu^- \tr{\chi^+}$  &  $ u^{Q,+s}_{3,1} \gamma_5 D_\mu^+ D_\nu^-  \tr{\chi^+}$  \\
			      &2 & $u^{Q,-}_{3,2} \gamma_5 D_\mu^+ D_\nu^- D^+ \! \! \cdot \! D^+$  &  $ u^{Q,+s}_{3,2}  \gamma_5 D_\mu^+ D_\nu^-  D^+ \! \! \cdot \! D^+$  \\
			      &3 &$u^{Q,+s}_{3,3} \sigma_\mu^{\ \alpha} D_\alpha^+ i D_\nu^- \tr{\chi^+}$ & $u^{Q,-}_{3,3} \sigma_\mu^{\ \alpha} D_\alpha^+ i D_\nu^- \tr{\chi^+}$  \\
			      &4 &$u^{Q,+s}_{3,4} \sigma_\mu^{\ \alpha} D_\alpha^+ i D_\nu^- D^+ \! \! \cdot \! D^+$ & $u^{Q,-}_{3,4} \sigma_\mu^{\ \alpha} D_\alpha^+ i D_\nu^- D^+ \! \! \cdot \! D^+$		      
\end{tabular}
\end{ruledtabular}
\end{table}%
where $\Gamma \in \{ \mathds{1},\gamma_5,\gamma_\mu,\gamma_\mu \gamma_5,\sigma_{\mu \nu},\sigma_{\mu \nu} \gamma_5\}$. These relations also hold if we insert additional chiral building blocks or an arbitrary number of derivatives. They are similar to those given in Refs.~\cite{Fettes:2000gb} and~\cite{Baur:1998rn}. For the different possible Clifford matrices we find explicitly
\begin{subequations}
\begin{flalign}
&\underline{\mathds{1}} & \bar{\Psi} \gamma^\beta D^+_\beta \Psi &\overset{\cdot}{=}  0 \ , \label{EOM_1a} \\*
& & \bar{\Psi} \gamma^\beta D^-_\beta \Psi &\overset{\cdot}{=} - 2 i m \bar{\Psi} \Psi \ , \label{EOM_1b}  \\[.5\baselineskip]
&\underline{\gamma_5} & \bar{\Psi} \gamma^\beta \gamma_5 D^-_\beta \Psi &\overset{\cdot}{=}  0 \ , \label{EOM_5a} \\*
& & \bar{\Psi} \gamma^\beta \gamma_5 D^+_\beta \Psi &\overset{\cdot}{=} - 2 i m \bar{\Psi} \gamma_5 \Psi \ , \label{EOM_5b}  \\[.5\baselineskip]
&\underline{\gamma^\mu} & \bar{\Psi} i \sigma^{\mu \beta} D^-_\beta \Psi &\overset{\cdot}{=} \bar{\Psi} D^{+\mu} \Psi \ , \label{EOM_mua} \\*
& & \bar{\Psi} i \sigma^{\mu \beta} D^+_\beta \Psi &\overset{\cdot}{=}  \bar{\Psi} D^{-\mu} \Psi + 2 i m \bar{\Psi} \gamma^\mu \Psi \ , \label{EOM_mub}  \\[.5\baselineskip]
&\underline{\gamma^\mu \gamma_5} & \bar{\Psi} \frac{1}{2} \epsilon^{\mu\beta\rho\sigma} \sigma_{\rho \sigma} D^+_\beta \Psi &\overset{\cdot}{=} \bar{\Psi} \gamma_5 D^{-\mu} \Psi \ , \label{EOM_mu5a} \\*
& & \bar{\Psi} \frac{1}{2} \epsilon^{\mu\beta\rho\sigma} \sigma_{\rho \sigma} D^-_\beta \Psi &\overset{\cdot}{=}  \bar{\Psi} \gamma_5 D^{+\mu} \Psi - 2 i m \bar{\Psi} \gamma^\mu \gamma_5 \Psi \ , \label{EOM_mu5b}  \\[.5\baselineskip]
&\underline{\sigma^{\mu \nu}}  & \bar{\Psi} \epsilon^{\mu\nu\beta\delta} \gamma_\delta \gamma_5 D^+_\beta \Psi &\overset{\cdot}{=} -i \bar{\Psi} \bigl(\gamma^\mu D^{-\nu} - \gamma^\nu D^{-\mu} \bigr) \Psi \label{EOM_sigmaa} \ , \\*
& & \bar{\Psi} \epsilon^{\mu\nu\beta\delta} \gamma_\delta \gamma_5 D^-_\beta \Psi &\overset{\cdot}{=}  - i \bar{\Psi}  \bigl(\gamma^\mu D^{+\nu} - \gamma^\nu D^{+\mu} \bigr) \Psi \nonumber \\*
& & &\quad- 2 i m \bar{\Psi} \sigma^{\mu \nu} \Psi \ , 
\label{EOM_sigmab}  \\[.5\baselineskip]
&\underline{\sigma^{\mu \nu} \gamma_5} \!\! &  \bar{\Psi} \epsilon^{\mu\nu\beta\delta} \gamma_\delta D^-_\beta \Psi &\overset{\cdot}{=} -i \bar{\Psi} \bigl(\gamma^\mu \gamma_5 D^{+\nu} - \gamma^\nu \gamma_5 D^{+\mu} \bigr) \Psi \ , \label{EOM_sigma5a} \\*
& & \bar{\Psi} \epsilon^{\mu\nu\beta\delta} \gamma_\delta D^+_\beta \Psi &\overset{\cdot}{=}  - i \bar{\Psi}  \bigl(\gamma^\mu \gamma_5 D^{-\nu} - \gamma^\nu \gamma_5 D^{-\mu} \bigr) \Psi \nonumber \\*
&&&\quad+ m \bar{\Psi} \epsilon^{\mu\nu\rho\sigma} \sigma_{\rho \sigma} \Psi \ . \label{EOM_sigma5b}
\end{flalign}
\end{subequations}
Note that, owing to the identity $\sigma^{\mu\nu} \gamma_5 = \frac{i}{2} \epsilon^{\mu\nu\rho\sigma} \sigma_{\rho\sigma}$, Eqs.~\eqref{EOM_sigma5a} and~\eqref{EOM_sigma5b} are actually only useful reformulations of Eqs.~\eqref{EOM_sigmaa} and~\eqref{EOM_sigmab}. We use Eqs.~\eqref{EOM_1a}-\eqref{EOM_mua} to neglect all terms where a derivative is contracted with an element of the Clifford algebra apart from $\sigma^{\mu \beta} D^+_\beta$. We use Eq.~\eqref{EOM_mub}, which is the Gordon identity, to replace all terms that simultaneously have the identity matrix as Dirac structure and contain a minus derivative. The remaining Eqs.~\eqref{EOM_mu5a}-\eqref{EOM_sigma5b} are used to dispose of as many structures containing the $\epsilon$ tensor as possible. Another useful identity is the curvature relation,
\begin{align} \label{curvature_relation}
 [D_\mu,D_\nu]=\frac{1}{4}[u_\mu,u_\nu] \ ,
\end{align}
which also holds for derivatives acting to the left and allows us to consider only symmetric combinations of derivatives. The EOM yields three additional relations,
\begin{subequations}
\begin{align}
 \bar{\Psi} \Gamma D^+ \! \! \cdot \! D^- \Psi &= \bar{\Psi} \bigl( \Gamma \slashed{\Dr}\slashed{\Dr} - \slashed{\Dl}\slashed{\Dl} \Gamma \bigr) \Psi \overset{\cdot}{=} 0 \ , \\
\begin{split}
  \bar{\Psi} \Gamma D^\pm \! \! \cdot \! D^\pm \Psi &= \bar{\Psi} \bigl( \Gamma \slashed{\Dr}\slashed{\Dr} + \slashed{\Dl}\slashed{\Dl}\Gamma \pm \Dl^{\vshiftindex{\alpha}{-4pt}} \Gamma \Dr_\alpha \bigr) \Psi \\ 
  &\overset{\cdot}{=} \bar{\Psi} \bigl(-2 m^2 \Gamma \pm \Dl^{\vshiftindex{\alpha}{-4pt}} \Gamma \Dr_\alpha \bigr) \Psi \ ,                                                                                                                                                                                                                                                                                                              
\end{split}
\end{align}
\end{subequations}
which we use to neglect all contractions of derivatives (acting on nucleon fields) apart from $D^+ \! \! \cdot \! D^+$. Rewriting the flavor matrix,
\begin{align}
 Q = Q - \frac{1}{2} \tr{Q} +  \frac{1}{2} \tr{Q} \equiv \tilde{Q} + \frac{1}{2} \tr{Q} , 
\end{align}
one finds that the trace part either vanishes or can (by a redefinition of LECs) be absorbed into a singlet contribution. We write down the low-energy form of $\mathcal{O}$ and $\tilde{\mathcal{O}}$ as
\begin{subequations}
\begin{align}
\mathcal{O}_{\mu\nu}^Q \Bigr|_{N \pi, n} &= \mathbf{S} \sum_i \bar{\Psi} \mathcal{O}^{n,i}_{\mu\nu} \Psi \ , \\
\tilde{\mathcal{O}}_{\mu\nu}^Q \Bigr|_{N \pi, n} &= \mathbf{S} \sum_i \bar{\Psi} \tilde{\mathcal{O}}^{n,i}_{\mu\nu}  \Psi \ ,
\end{align}
\end{subequations}
where $n$ indicates the chiral order. The $\mathcal{O}^{n,i}_{\mu\nu}$ and $\tilde{\mathcal{O}}^{n,i}_{\mu\nu}$, given in Table~\ref{LE_Operators}, can be expressed economically with the following abbreviations for recurring structures,
\begin{subequations}
 \begin{align}
 u_{i,j}^{Q,\pm} &= l_{i,j} \bigl( u^\dagger \tilde{Q} u \pm u \tilde{Q} u^\dagger \bigr) \ , \\
 u_{i,j}^{Q,s} &= 2 l_{i,j}^s \tr{Q} \ , \\
 u_{i,j}^{Q,+s} &= u_{i,j}^{Q,+}+ u_{i,j}^{Q,s} \ ,
\end{align}
\end{subequations}
where the $l_{i,j}$ and $l_{i,j}^s$ are LECs.

%
%

\section{Calculation \label{sect_calculation}}
\begin{figure}[tb]
\subfigure[\label{fd:subfiga}]{
\includegraphics[scale=0.55]{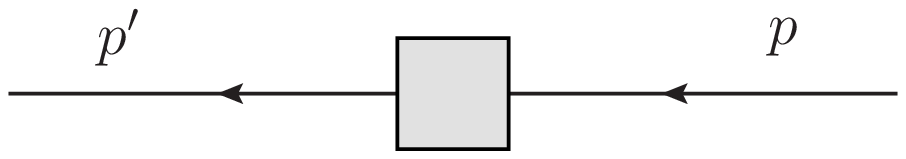}
}
\subfigure[\label{fd:subfigb}]{
\includegraphics[scale=0.55]{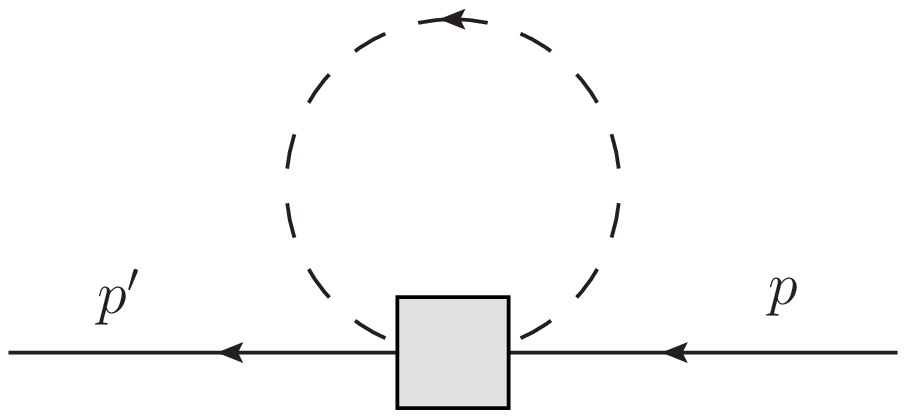}
}
\subfigure[\label{fd:subfigc}]{
\includegraphics[scale=0.55]{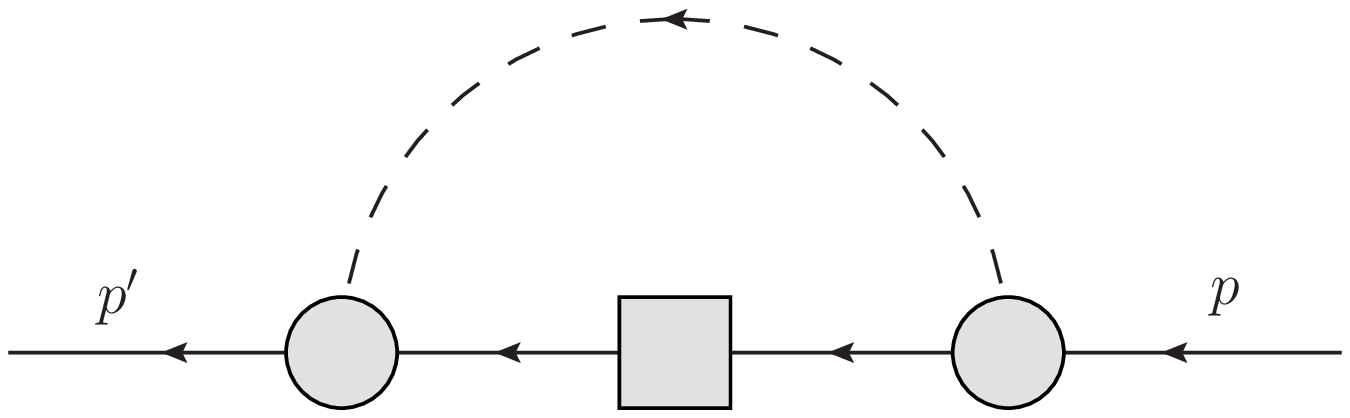}
}
\subfigure[\label{fd:subfigd}]{
\includegraphics[scale=0.55]{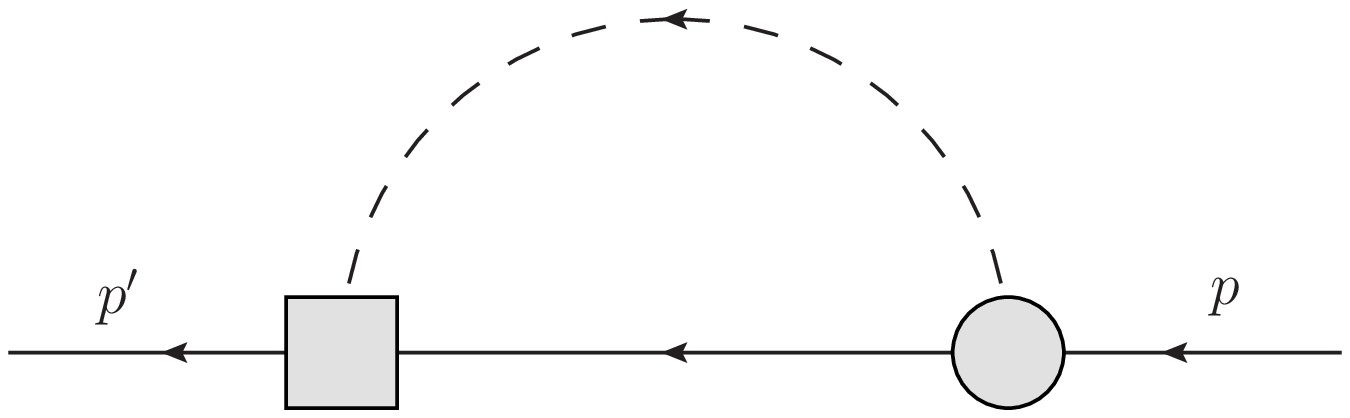}
}
\subfigure[\label{fd:subfige}]{
\includegraphics[scale=0.55]{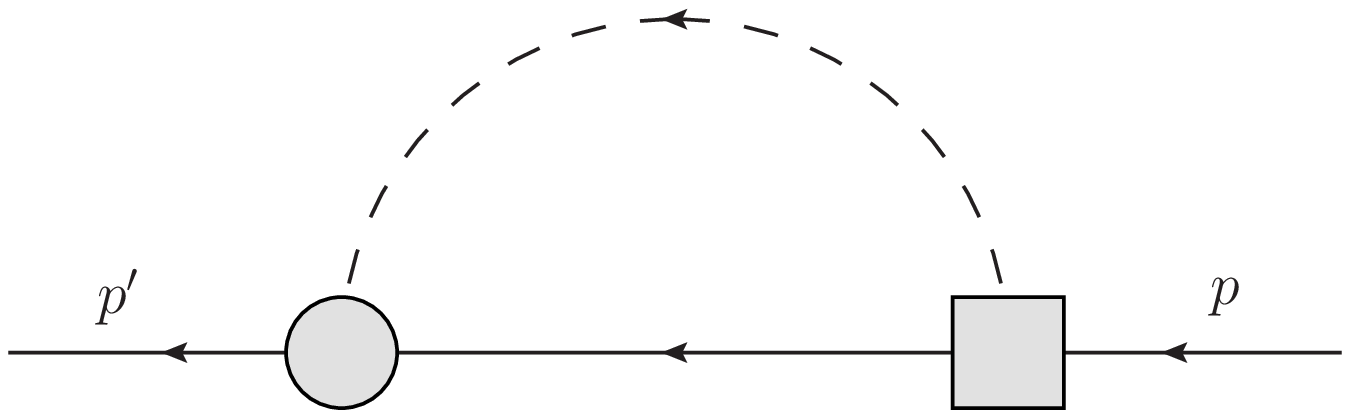}
}
\subfigure[\label{fd:subfigf}]{
\includegraphics[scale=0.55]{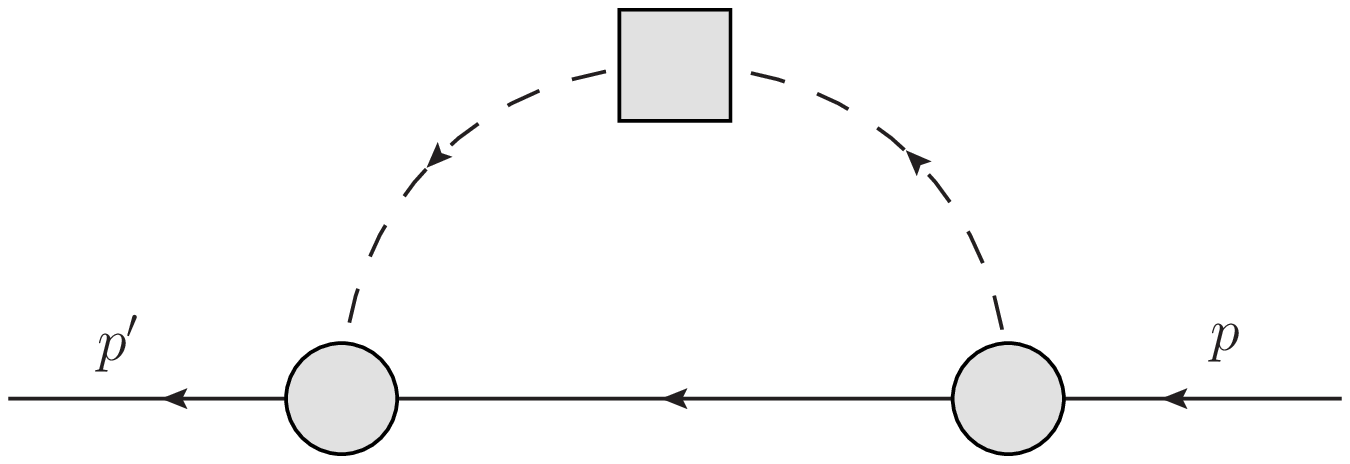}
}
\caption[]{\label{FD_nGPDs} Feynman diagrams needed for the full one-loop calculation. Diagram \subref{fd:subfigf} only contributes to the (isosinglet) vector current. Diagram \subref{fd:subfiga} has to be multiplied with a $Z$ factor if the operator insertion is of zeroth or first order. Additional Feynman diagrams containing quark mass insertions from the second-order pion-nucleon Lagrangian are not shown explicitly.}
\end{figure}%
The different types of Feynman diagrams needed for the full one-loop calculation in BChPT of the first moments of the quark GPDs are shown in Fig.~\ref{FD_nGPDs}. Full one-loop in this context means that we take into account all orders of ChPT that do not contain two-loop contributions (i.e.\ up to and including third order). Diagram type \subref{fd:subfiga} starts to contribute at zeroth, \subref{fd:subfigb}-\subref{fd:subfige} at second and \subref{fd:subfigf} at third order. Solid (dashed) lines depict nucleon (meson) propagators. The circles stand for vertices from the chiral Lagrangian while the squares indicate the operator insertion. Additional diagrams with quark mass insertions from the second-order pion-nucleon Lagrangian are not depicted in Fig.~\ref{FD_nGPDs}. We take them into account via replacing $m_0$ by $m_N=m_0 - 4 c_1 m_\pi^2$ everywhere. Due to this procedure our results sporadically contain higher-order diagrams. \par
In order to present the results in the most convenient way we utilize the form factor decomposition given in Eq.~\eqref{formfactordecomposition} and use the isoscalar and isovector combinations of the generalized form factors. They are defined as
\begin{subequations}
\begin{align}
 A^{s,v}_{2,0}&=\bigl(A^u_{2,0} \pm A^d_{2,0}\bigr)^p = \pm \bigl(A^u_{2,0} \pm A^d_{2,0}\bigr)^n \ , \\
 B^{s,v}_{2,0}&=\bigl(B^u_{2,0} \pm B^d_{2,0}\bigr)^p = \pm \bigl(B^u_{2,0} \pm B^d_{2,0}\bigr)^n \ , \\
 C^{s,v}_{2}&=\bigl(C^u_{2} \pm C^d_{2}\bigr)^p = \pm \bigl(C^u_{2} \pm C^d_{2}\bigr)^n \ , \\
 \tilde{A}^{s,v}_{2,0}&=\bigl(\tilde{A}^u_{2,0} \pm \tilde{A}^d_{2,0}\bigr)^p = \pm \bigl(\tilde{A}^u_{2,0} \pm \tilde{A}^d_{2,0}\bigr)^n \ , \\
 \tilde{B}^{s,v}_{2,0}&=\bigl(\tilde{B}^u_{2,0} \pm \tilde{B}^d_{2,0}\bigr)^p = \pm \bigl(\tilde{B}^u_{2,0} \pm \tilde{B}^d_{2,0}\bigr)^n \ , 
\end{align}
\end{subequations}
where the superscripts $p$ and $n$ are introduced to differentiate between proton and neutron form factors. For these form factors we obtain explicit expressions in terms of elementary functions and a remaining integration over a Feynman parameter (originating from the three point functions contained in diagrams of type \subref{fd:subfigc} and \subref{fd:subfigf}). In general nucleon four-momenta have to be counted as large (zeroth order) in BChPT. However $\Delta=p^\prime-p$ has to be counted as small (first order) since the nucleon mass drops out in the momentum difference. This is also true for the $\Delta$'s in the decomposition of the matrix elements (see Eq.~\eqref{formfactordecomposition}) and, accordingly, our results for $A^{s,v}_{2,0}$, $\tilde{A}^{s,v}_{2,0}$ are exact to third, those for $B^{s,v}_{2,0}$, $\tilde{B}^{s,v}_{2,0}$ to second and the one for $C^{s,v}_{2}$ to first chiral order.\par
The full one-loop result for the generalized form factors obtained by a manifestly covariant calculation in the infrared regularization scheme \cite{Becher:1999he} is presented in Appendix~\ref{app_full_result}. Up to the order to which our results are exact the occurring divergences can be absorbed in LECs by the use of Eq.~\eqref{renormalized_LECs}. The higher-order divergences have to be canceled by hand which introduces the unphysical scale dependence in higher-order terms typical for manifestly covariant calculations. In order to obtain a meaningful result one has to set this scale to a typical hadronic value around $\unit{1}{\giga\electronvolt}$. A variation of this scale within reasonable bounds, say $\unit{0.8}{\giga\electronvolt}$-$\unit{1.2}{\giga\electronvolt}$, can (and should) be used to estimate the systematic error due to higher-order effects (cf.~\cite{Lacour:2007wm}, where the $\rho$ ($\Xi$) mass has been chosen as lower (upper) bound).\par
Taking a closer look at the results one finds that many new LECs occur. They mainly originate from first-order operator insertions in diagrams of type \subref{fd:subfigd} and \subref{fd:subfige}. While at least in principle one would prefer to keep all these structures, the reality of a finite amount of available lattice data might force us to reduce the number of free parameters in the end. One possibility to achieve such a reduction would be a strict truncation of the results at the order to which they are exact, which leads to the same extrapolation formulas as an equivalent calculation within the framework of heavy BChPT (cf.\ Sec.~\ref{sect_heavy_baryon_reduction}). Doing so one would however loose all benefits of the covariant result, like the correct behavior near the two-pion threshold at $t=4 m_\pi^2$ and the improved convergence that is often ascribed to the untruncated results. A reasonable compromise could look as follows: Truncate diagrams \subref{fd:subfigd} and \subref{fd:subfige}, which yield the main share of the new LECs, but keep the triangle diagrams \subref{fd:subfigc} and \subref{fd:subfigf}, which do not introduce new LECs, to all orders. This tradeoff is particularly appealing since diagram \subref{fd:subfigf} is responsible for the threshold behavior that is not described correctly by the truncated version. \par
The low-energy versions of the gluon operator $\mathcal{O}^g_{\mu\nu}$ and the singlet combination of the quark operator $\mathcal{O}^\mathds{1}_{\mu\nu}$ are equal up to the numerical value of the LECs since they behave similar under chiral rotations, charge conjugation and parity transformation. We therefore define the operator by a mere replacement of LECs:
\begin{align}
\mathcal{O}^g_{\mu\nu} \equiv \mathcal{O}^\mathds{1}_{\mu\nu}\quad (l^s \rightarrow l^g, l^s_{i,j} \rightarrow l^g_{i,j}) \ .
\end{align}
The generalized form factors inherit this property and are obtained via replacing all superscripts $s$ by $g$. 

\section{Heavy baryon reduction \label{sect_heavy_baryon_reduction}}

To obtain the heavy baryon reduced result one has to truncate the simultaneous expansion in the pion mass $m_\pi$ and the momentum transfer $t=\Delta^2$ at full one-loop order. To be consistent with the chiral counting scheme one has to count $t$ as second order. This can be achieved most easily by keeping the ratio $t/m_\pi^2$, which is counted as $\mathcal{O}(1)$, fixed.

\subsection{Generalized form factors}
The heavy baryon reduced version of our result has the advantage that it can be written in a compact and lucid way. Writing all functions in such a way that only real quantities occur in the spacelike region, we obtain for the form factors:
\begin{subequations} \label{result_GFF_nucleon}
\begin{align}
\begin{split}
 A^s_{2,0}&=A_{2,0}^{s,(0)}+ A_{2,0}^{s,(m2)} m_\pi^2 + A_{2,0}^{s,(m3)} m_\pi^3+A_{2,0}^{s,(t)} t\\
&\quad+\frac{3 \pi A_{2,0}^{\pi,s,(0)} g_A^2 \left(8 m_\pi^4-6 m_\pi^2 t+t^2\right) }{16 m_0 (4 \pi F_\pi)^2 \sqrt{-t}} \arcsin\left(\frac{1}{\sigma }\right)\\
&\quad-\frac{3 \pi A_{2,0}^{\pi,s,(0)} g_A^2 }{8 m_0 (4 \pi F_\pi)^2 } m_\pi t + \mathcal{O}(p^4) \ ,
\end{split} \raisetag{40pt} \\[8pt]
\begin{split}
B^s_{2,0}&=B_{2,0}^{s,(0)}+B_{2,0}^{s,(m2)}(\mu) m_\pi^2 +B_{2,0}^{s,(t)}(\mu)t\\
&\quad- 3 g_A^2 \frac{A_{2,0}^{s,(0)}+B_{2,0}^{s,(0)}-A_{2,0}^{\pi,s,(0)}}{(4 \pi F_\pi)^2} m_\pi^2 \ln\frac{m_\pi^2}{\mu^2} \\
&\quad -  \frac{A_{2,0}^{\pi,s,(0)}g_A^2 }{2 (4 \pi F_\pi)^2} t \ln\frac{m_\pi^2}{\mu^2} \\
&\quad-\frac{A_{2,0}^{\pi,s,(0)} g_A^2 }{(4 \pi F_\pi)^2} t \sigma^3 \arctanh\left(\frac{1}{\sigma }\right)  + \mathcal{O}(p^3) \ ,
\end{split}  \\[8pt]
\begin{split}
 C^s_{2}&=C_{2}^{s,(0)}+\frac{3 \pi A_{2,0}^{\pi,s,(0)} g_A^2 m_0 m_\pi \left(-2 m_\pi^2+t\right)}{8 (4 \pi F_\pi)^2  t}\\
&\quad+\frac{3\pi A_{2,0}^{\pi,s,(0)} g_A^2 m_0 \left(-8 m_\pi^4+2 m_\pi^2 t+t^2\right)}{16 (4 \pi F_\pi)^2 (-t)^{3/2}} \arcsin\left(\frac{1}{\sigma }\right)\\
&\quad + \mathcal{O}(p^2)\ ,
\end{split} \raisetag{10pt} \\[8pt]
\begin{split}
 \tilde{A}^s_{2,0}&= \tilde{A}_{2,0}^{s,(0)}+\tilde{A}_{2,0}^{s,(m2)}(\mu)m_\pi^2 +\tilde{A}_{2,0}^{s,(m3)} m_\pi^3+\tilde{A}_{2,0}^{s,(t)} t\\
&\quad-\frac{3 \tilde{A}_{2,0}^{s,(0)} g_A^2 }{(4 \pi F_\pi)^2} m_\pi^2 \ln\frac{m_\pi^2}{\mu^2} + \mathcal{O}(p^4) \ ,
\end{split} \raisetag{18pt} \\[8pt]
\begin{split}
\tilde{B}^s_{2,0}&= \tilde{B}_{2,0}^{s,(0)} +\tilde{B}_{2,0}^{s,(m2)}(\mu) m_\pi^2 +\tilde{B}_{2,0}^{s,(t)} t\\
&\quad-\frac{(\tilde{A}_{2,0}^{s,(0)}+3 \tilde{B}_{2,0}^{s,(0)}) g_A^2 }{(4 \pi F_\pi)^2} m_\pi^2 \ln\frac{m_\pi^2}{\mu^2} + \mathcal{O}(p^3) \ ,
\end{split}  \\[8pt]
\begin{split}
 A^v_{2,0}&=  A_{2,0}^{v,(0)}+ A_{2,0}^{v,(m2)}(\mu) m_\pi^2+A_{2,0}^{v,(m3)} m_\pi^3+A_{2,0}^{v,(t)} t\\
&\quad-\frac{A_{2,0}^{v,(0)} \left(1+3 g_A^2\right) }{(4 \pi F_\pi)^2}m_\pi^2 \ln\frac{m_\pi^2}{\mu^2} + \mathcal{O}(p^4) \ ,
\end{split} \raisetag{18pt}  \\[8pt]
\begin{split}
B^v_{2,0}&=B_{2,0}^{v,(0)}+B_{2,0}^{v,(m2)}(\mu) m_\pi^2 +B_{2,0}^{v,(t)}t\\
&\quad-\frac{B_{2,0}^{v,(0)}-(A_{2,0}^{v,(0)}-2 B_{2,0}^{v,(0)}) g_A^2 }{(4 \pi F_\pi)^2} m_\pi^2 \ln\frac{m_\pi^2}{\mu^2}\\
&\quad + \mathcal{O}(p^3) \ ,
\end{split}  \\[8pt]
\begin{split} \label{result_GFF_nucleon_Cv2}
C^v_{2}&=C_{2}^{v,(0)} + \mathcal{O}(p^2)\ ,
\end{split}  \\[8pt]
\begin{split}
 \tilde{A}^v_{2,0}&=\tilde{A}_{2,0}^{v,(0)}+ \tilde{A}_{2,0}^{v,(m2)}(\mu)m_\pi^2+\tilde{A}_{2,0}^{v,(m3)} m_\pi^3+\tilde{A}_{2,0}^{v,(t)} t\\
&\quad-\frac{\tilde{A}_{2,0}^{v,(0)} \left(1+2 g_A^2\right) }{(4 \pi F_\pi)^2} m_\pi^2 \ln\frac{m_\pi^2}{\mu^2}+ \mathcal{O}(p^4) \ ,
\end{split} \raisetag{18pt}  \\[8pt]
\begin{split}
\tilde{B}^v_{2,0}&=\tilde{B}_{2,0}^{v,(0)} +\tilde{B}_{2,0}^{v,(m2)}(\mu)m_\pi^2 +\tilde{B}_{2,0}^{v,(t)} t \\
&\quad+\frac{(\tilde{A}_{2,0}^{v,(0)}-6 \tilde{B}_{2,0}^{v,(0)}) g_A^2-3 \tilde{B}_{2,0}^{v,(0)}}{3 (4 \pi F_\pi)^2} m_\pi^2 \ln\frac{m_\pi^2}{\mu^2}\\
&\quad + \mathcal{O}(p^3) \ ,
\end{split}
\end{align}
\end{subequations}
where
\begin{align}
 \sigma = \sqrt{1-\frac{4 m_\pi^2}{t}} \ ,
\end{align}
and the fit parameters are related to the original LECs defined in Sec.~\ref{sect_operator_construction} by Eqs.~\eqref{def_fit_parameters_isoscalar} and~\eqref{def_fit_parameters_isovector}. The fit parameter $A^{\pi,s,(0)}_{2,0}$ originating from diagram \subref{fd:subfigf}, where the operator couples to two pions, should not be treated as completely free. Instead one should obtain it from a fit to the pion GPD, where it fixes the chiral limit value in the forward case. For consistency we propose to use the one-loop result derived in Ref.~\cite{Diehl:2005rn} (see also Refs.~\cite{Donoghue:1991qv,Kivel:2002ia}), that reads
\begin{subequations} \label{result_GFF_pion}
\begin{align}
A^{\pi,s}_{2,0}&=A^{\pi,s,(0)}_{2,0} + A^{\pi,s,(m2)}_{2,0} m_\pi^2 + A^{\pi,s,(t)}_{2,0} t + \mathcal{O}(p^4) \ , \\[8pt]
\begin{split}
A^{\pi,s}_{2,2}&=-\frac{1}{4} A^{\pi,s,(0)}_{2,0} + A^{\pi,s,(m2)}_{2,2}(\mu) m_\pi^2 + A^{\pi,s,(t)}_{2,2}(\mu) t \\
&\quad-\frac{1}{4} A^{\pi,s,(0)}_{2,0} \frac{m_\pi^2-2t}{3 (4 \pi F_\pi)^2} \biggl(\ln\frac{m_\pi^2}{\mu^2}+\frac{4}{3} - \frac{t+2m_\pi^2}{t} J(t) \biggr) \\
&\quad  + \mathcal{O}(p^4) \ ,
\end{split} \raisetag{10pt} 
\end{align}
\end{subequations}
where
\begin{align}
 J(t)= 2 + \sigma \ln \frac{\sigma-1}{\sigma+1} \ .
\end{align}
In Ref.~\cite{Diehl:2005rn} all moments of GPDs have been treated simultaneously by the use of nonlocal operators, which is obviously a very elegant method. For the specific case of the first moments, connected to the form factors shown above, we were able to confirm this result by a straight forward calculation with local operators.
\subsection{Value and slope in the forward limit}
For small values of the momentum transfer $|t| \ll 4 m_\pi^2$ the form factors are often represented by a Taylor expansion in $t$. We use the notation of Ref.~\cite{Dorati:2007bk},
\begin{align}
 X^{s,v}(t) &= X^{s,v}(0) + \rho^{s,v}_X t + \mathcal{O}(t^2) \ ,
\end{align}
where $X$ can stand for arbitrary form factors. For the nontrivial cases one obtains
\begin{subequations} \label{result_GFF_nucleon_forward_limit}
\begin{align}
\begin{split}  
 A^s_{2,0}(0) &= A_{2,0}^{s,(0)}+ A_{2,0}^{s,(m2)} m_\pi^2\\
  &\quad  + \biggl(A_{2,0}^{s,(m3)} + \frac{3 \pi A^{\pi,s,(0)}_{2,0} g_A^2}{4 m_0 (4 \pi F_\pi)^2} \biggr) m_\pi^3 + \mathcal{O}(p^4) \ ,
\end{split} \raisetag{18pt}  \\[8pt]
\begin{split}  
 B^s_{2,0}(0) &= B_{2,0}^{s,(0)}+\biggl(B_{2,0}^{s,(m2)}(\mu)+\frac{4 A^{\pi,s,(0)}_{2,0} g_A^2}{(4 \pi F_\pi)^2} \biggr) m_\pi^2 \\
  &\quad  - 3 g_A^2 \frac{A_{2,0}^{s,(0)}+B_{2,0}^{s,(0)}-A_{2,0}^{\pi,s,(0)}}{(4 \pi F_\pi)^2} m_\pi^2 \ln\frac{m_\pi^2}{\mu^2} \\
&\quad+  \mathcal{O}(p^3) \ ,
\end{split}  \raisetag{10pt}  \\[8pt]
C^s_2(0) &= C_2^{s,(0)} + \frac{ \pi A^{\pi,s,(0)}_{2,0} g_A^2 m_0}{4 (4 \pi F_\pi)^2} m_\pi  + \mathcal{O}(p^2)\ , \\[8pt]
\begin{split}  
A^{\pi,s}_{2,2}(0)&=-\frac{1}{4} A^{\pi,s,(0)}_{2,0} + \biggl(A^{\pi,s,(m2)}_{2,2}(\mu) - \frac{A^{\pi,s,(0)}_{2,0}}{12 (4 \pi F_\pi)^2} \biggr) m_\pi^2\\
&\quad-\frac{A^{\pi,s,(0)}_{2,0}}{12 (4 \pi F_\pi)^2} m_\pi^2 \ln \frac{m_\pi^2 }{\mu^2} + \mathcal{O}(p^4) \ , \raisetag{18pt}
\end{split}
\end{align}
\end{subequations}
and
\begin{subequations} \label{result_slope_GFF_nucleon_forward_limit}
\begin{align}
  \rho^s_{A_{2,0}} &= A_{2,0}^{s,(t)}  - \frac{7 \pi A^{\pi,s,(0)}_{2,0} g_A^2}{8 m_0 (4 \pi F_\pi)^2} m_\pi + \mathcal{O}(p^2) \ ,\\[8pt]
\begin{split}
\rho^s_{B_{2,0}}&= B_{2,0}^{s,(t)}(\mu) - \frac{4 A^{\pi,s,(0)}_{2,0} g_A^2}{3 (4 \pi F_\pi)^2}\\
&\quad - \frac{A_{2,0}^{\pi,s,(0)}g_A^2 }{2 (4 \pi F_\pi)^2} \ln\frac{m_\pi^2}{\mu^2} + \mathcal{O}(p^1) \ ,
\end{split}\\[8pt]
 \rho^s_{C_2} &= - \frac{\pi A^{\pi,s,(0)}_{2,0} g_A^2 m_0}{10 (4 \pi F_\pi)^2} m_\pi^{-1} + \mathcal{O}(p^0) \ , \\[8pt]
\begin{split}  
\rho^s_{A^\pi_{2,2}} &= A^{\pi,s,(t)}_{2,2}(\mu)  + \frac{11 A^{\pi,s,(0)}_{2,0}}{60 (4 \pi F_\pi)^2} + \frac{A^{\pi,s,(0)}_{2,0}}{6(4 \pi F_\pi)^2} \ln \frac{m_\pi^2}{\mu^2}\\
&\quad + \mathcal{O}(p^2) \ . \raisetag{10pt}
\end{split}
\end{align}
\end{subequations}
The formula for the slope is always two orders less accurate than the one for the corresponding form factor since one has to take a derivative with respect to $t$ to obtain it. The forward limit values and slopes of the form factors not given here explicitly can easily be obtained from Eqs.~\eqref{result_GFF_nucleon} and~\eqref{result_GFF_pion}.\par
Our result is consistent with the heavy baryon result provided in Refs.~\cite{Diehl:2006ya,Diehl:2006js}. In particular the chiral logarithms in the $B$-type form factors are reproduced, which was not yet the case for the leading one-loop calculation presented in Ref.~\cite{Dorati:2007bk}. Note that the truncated version of our result has the same accuracy as the above cited heavy baryon calculations for $B$-type form factors, but is one order more/less precise for the $A$/$C$-type ones. This is because in Refs.~\cite{Diehl:2006ya,Diehl:2006js}, contributions going beyond the one-loop level are taken into account for the $C$-type form factor. Note, however, that our untruncated covariant result already contains these contributions up to higher-order tree-level insertions, which could be added by hand if necessary.

\section{Summary and outlook}

In this work we have extended the results of Ref.~\cite{Dorati:2007bk} to full one-loop accuracy and provide chiral extrapolation formulas for the first moments of chiral-even nucleon GPDs. Due to the overlap in LECs some of the form factors should be fit to lattice data simultaneously. If one uses the heavy baryon reduced fit formulas, those are the two pairs $(A^v_{2,0},B^v_{2,0})$ and $(\tilde{A}^v_{2,0},\tilde{B}^v_{2,0})$ in the isotriplet sector. In the isosinglet sector the situation is more entangled and calls for simultaneous fits of $(A^s_{2,0},B^s_{2,0},C^s_2,A^{\pi,s}_{2,0},A^{\pi,s}_{2,2})$ and $(\tilde{A}^s_{2,0},\tilde{B}^s_{2,0})$. If the reader is more ambitious and wants to apply the untruncated formulas given in Appendix~\ref{app_full_result}, he has to treat all form factors from the isosinglet/isotriplet sector simultaneously. In the ideal case one has also data for the matrix elements of the gluonic operators: in a first step one would fit the two sectors separately and check whether the sum rules given in Eq.~\eqref{sum_rules} are fulfilled to a satisfactory degree. Afterwards one can invoke the sum rules to restrict the LECs in order to find the most precise result.\par
It is reasonable to expect the range of applicability of covariant BChPT to extend to $m_{\pi,\text{max}} \approx \unit{300}{\mega\electronvolt}$. One can a priori not say for sure how far our formulas are applicable in $t$ direction. A careful estimate based on the argument that the chiral counting scheme ranks $t$ as a quantity of second order leads to $|t_{\text{max}}| \sim m_{\pi,\text{max}}^2 \approx \unit{0.1}{\giga\squaren\electronvolt}$. For simulations with periodic boundary conditions one will need rather large volumes to find one ($L\approx\unit{4}{\femto\metre}$) or two ($L\approx\unit{5.5}{\femto\metre}$) nonzero momentum configurations in this region. It is therefore quite unfortunate that the $B$- and $C$-type form factors are only accessible at nonzero momentum transfer (compare Eq.~\eqref{formfactordecomposition}) and that our extrapolation formulas for them are only accurate to second (first) chiral order. To reach high accuracy one therefore presumably needs simulations with twisted boundary conditions, where low momentum transfer is accessible without going to very large volumes.\par
One can think of several extensions of this work. For instance, a three-flavor calculation at full one-loop level, extending the work done in Ref.~\cite{Bruns:2011sh}, seems fruitful, since lattice simulations of three-point functions with $2+1$ dynamical quark flavors are in progress. Other possibilities are the investigation of (possibly relevant) decuplet contributions, isospin-breaking and finite volume effects~\cite{Detmold:2005pt}. Another challenging topic would be the (simultaneous) calculation of all $x$ moments by the use of nonlocal operators. In Ref.~\cite{Moiseeva:2012zi} it has been shown that such an approach is only applicable in a covariant framework and that particular care has to be taken of the regularization procedure. In our opinion this interesting observation calls for further investigation.

\appendix

\section{Low-energy constants and fit parameters \label{app_LECs}}
The divergences occurring in the loop calculation for $4-d=\epsilon \rightarrow 0$ have to be absorbed in LECs,
\begin{align} \label{renormalized_LECs}
 l_{i,j}&= l^{(r)}_{i,j}(\mu) + \gamma_{i,j} L \ , &  l^s_{i,j}&= l^{s,(r)}_{i,j}(\mu) + \gamma^s_{i,j} L  \ ,
\end{align}
where $L$ contains the pole term plus the typical constants for the modified minimal subtraction scheme,
\begin{align}
L &= \frac{-\mu^{-\epsilon}}{(4 \pi)^2} \left( \frac{1}{\epsilon}+\frac{1}{2} \left(1+\ln{4\pi}-\gamma_E \right) \right) \ .
\end{align}
The renormalized constants pick up a scale dependence,
\begin{align}
\mu \frac{\partial}{\partial \mu} l^{(r)}_{i,j}(\mu)&=\frac{-\gamma_{i,j}}{(4 \pi)^2} \ , & \mu \frac{\partial}{\partial \mu} l^{s,(r)}_{i,j}(\mu)&=\frac{-\gamma^s_{i,j}}{(4 \pi)^2} \ .
\end{align}
For the nonzero $\gamma_{i,j}$ and $\gamma^s_{i,j}$ we find
\begin{subequations}
\begin{align}  \label{nGPDs_scale_dependence_LECs}
\gamma_{2,2} & = \frac{1+3 g_A^2}{2 F_\pi^2} l_{0,1} \ ,  \\
\gamma_{2,4} & = \frac{1+2 g_A^2}{2 F_\pi^2} l_{0,2} \ ,  \\
\gamma_{3,1} & = \frac{1+2 g_A^2}{2 F_\pi^2} l_{1,1} - \frac{g_A^2}{12 m_0 F_\pi^2} l_{0,2} \ ,  \\
\gamma_{3,3} & = \frac{1+2 g_A^2}{2 F_\pi^2} l_{1,2} - \frac{g_A^2}{4 m_0 F_\pi^2} l_{0,1} \ ,  \\
\gamma^s_{2,4} & = \frac{3 g_A^2}{2 F_\pi^2} l^s_{0,2} \ , \\
\gamma^s_{3,1} & = \frac{g_A^2}{4 m_0 F_\pi^2} l^s_{0,2}+ \frac{3 g_A^2}{2 F_\pi^2} l^s_{1,1} \ , \\
\gamma^s_{3,3} & = \frac{3 g_A^2}{4 m_0 F_\pi^2} l^s_{0,1} +\frac{3 g_A^2}{2 F_\pi^2} l^s_{1,2}-\frac{3 g_A^2}{4 m_0 F_\pi^4} l^s \ , \\
\gamma^s_{3,4} & = \frac{-g_A^2}{ 2 F_\pi^4 m_0} l^s \ .
\end{align}
\end{subequations}
The combined fit parameters used in \eqref{result_GFF_nucleon} are related to the LECs defined in Sec.~\ref{sect_operator_construction} by
\begin{subequations} \label{def_fit_parameters_isoscalar}
\begin{align}
 A_{2,0}^{s,(0)}  &= 8 l^s_{0,1} \ , \\[2pt]
 A_{2,0}^{s,(m2)} &= 32 l^s_{2,2} \ , \\[2pt]
\begin{split}
 A_{2,0}^{s,(m3)} &= \frac{7 g_A^2 l^s}{8 F_\pi^4 m_0 \pi }-\frac{3 g_A^2 l^s_{0,1}}{4 F_\pi^2 m_0 \pi }-\frac{g_A l^s_{1,6}}{F_\pi^2 m_0 \pi }\\
&\quad -\frac{12 g_A m_0 l^{s}_{1,15}}{F_\pi^2 \pi }-\frac{12 g_A l^{s}_{1,18}}{F_\pi^2 \pi } \ ,
\end{split} \\[2pt]
 A_{2,0}^{s,(t)}  &= -8 l^s_{2,3} \ , \\[2pt]
 B_{2,0}^{s,(0)}        &= 16 m_0 l^s_{1,2} \ , \\[2pt]
 B_{2,0}^{s,(m2)}(\mu)  &= -\frac{2 g_A^2 l^{s}}{F_\pi^4 \pi^2}+ 64 m_0 l^{s,(r)}_{3,3}(\mu) -64 l^s_{1,2} c_1 \ , \\[2pt]
 B_{2,0}^{s,(t)}(\mu)   &= -16 m_0 l^{s,(r)}_{3,4}(\mu) + \frac{5 g_A^2 l^s}{12 F_\pi^4 \pi^2}\ , \\[8pt]
 C_{2}^{s,(0)}        &= -4 m_0 l^s_{2,1} \ , \\[2pt]
 \tilde{A}_{2,0}^{s,(0)}        &= 8 l^s_{0,2} \ , \\[2pt]
 \tilde{A}_{2,0}^{s,(m2)}(\mu)  &= -\frac{3 g_A^2 l^{s}_{0,2}}{2 F_\pi^2 \pi^2}+32 l^{s,(r)}_{2,4}(\mu) \ , \\[2pt]
 \tilde{A}_{2,0}^{s,(m3)}       &= \frac{7 g_A^2 l^{s}_{0,2}}{4 F_\pi^2 m_0 \pi }-\frac{g_A l^{s}_{1,4}}{F_\pi^2 m_0 \pi } \ , \\[2pt]
 \tilde{A}_{2,0}^{s,(t)}        &= -8 l^s_{2,5} \ , \\[2pt]
 \tilde{B}_{2,0}^{s,(0)}        &= 16 m_0 l^s_{1,1} \ , \\[2pt]
 \tilde{B}_{2,0}^{s,(m2)}(\mu)  &= -\frac{3 g_A^2 m_0 l^{s}_{1,1}}{F_\pi^2 \pi^2}+64 m_0 l^{s,(r)}_{3,1}(\mu)-64 l^s_{1,1} c_1 \ , \\[2pt]
 \tilde{B}_{2,0}^{s,(t)}        &= -16 m_0 l^{s}_{3,2} \ , \\[2pt]
 A^{\pi,s,(0)}_{2,0} &= \frac{8 l^s}{F_\pi^2} \ ,
\end{align}
\end{subequations}
and in the isovector sector by
\begin{subequations} \label{def_fit_parameters_isovector}
\begin{align} 
 A_{2,0}^{v,(0)}        &= 4 l_{0,1} \ , \\[2pt]
 A_{2,0}^{v,(m2)}(\mu)  &= -\frac{g_A^2 l_{0,1}}{2 F_\pi^2 \pi^2}+16 l^{(r)}_{2,2}(\mu) \ , \\[2pt]
\begin{split}
 A_{2,0}^{v,(m3)}       &= \frac{7 g_A^2 l_{0,1}}{8 F_\pi^2 m_0 \pi }+\frac{g_A l_{0,2}}{3 F_\pi^2 m_0 \pi }-\frac{g_A (l_{1,6}+l_{1,7})}{6 F_\pi^2 m_0 \pi }\\
&\quad -\frac{2 g_A m_0 (l_{1,15}+l_{1,16})}{F_\pi^2 \pi }-\frac{2 g_A (l_{1,18}+l_{1,19})}{F_\pi^2 \pi } \ ,
\end{split} \\[2pt]
 A_{2,0}^{v,(t)}        &= -4 l_{2,3} \ , \\[2pt]
 B_{2,0}^{v,(0)}        &= 8 m_0 l_{1,2} \ , \\[2pt]
 B_{2,0}^{v,(m2)}(\mu)  &= -\frac{g_A^2 m_0 l_{1,2}}{F_\pi^2 \pi^2} + 32 m_0 l^{(r)}_{3,3}(\mu)-32 l_{1,2} c_1 \ , \raisetag{18pt} \\[2pt]
 B_{2,0}^{v,(t)}        &= -8 m_0 l_{3,4}\ , \\[2pt]
 C_{2}^{v,(0)}        &= -2 m_0 l_{2,1} \ , \\[2pt]
\tilde{A}_{2,0}^{v,(0)}        &= 4 l_{0,2} \ , \\[2pt]
\tilde{A}_{2,0}^{v,(m2)}(\mu)  &= -\frac{g_A^2 l_{0,2}}{4 F_\pi^2 \pi^2}+16 l^{(r)}_{2,4}(\mu) \ , \\[2pt]
\begin{split}
 \tilde{A}_{2,0}^{v,(m3)}       &= \frac{g_A l_{0,1}}{3 F_\pi^2 m_0 \pi }+\frac{11 g_A^2 l_{0,2}}{24 F_\pi^2 m_0 \pi }+\frac{g_A l_{1,2}}{3 F_\pi^2 \pi }\\
&\quad +\frac{2 g_A l_{1,17}}{3 F_\pi^2 \pi }-\frac{g_A (l_{1,4}+l_{1,5})}{6 F_\pi^2 m_0 \pi } \ , 
\end{split} \\[2pt]
 \tilde{A}_{2,0}^{v,(t)}        &= -4 l_{2,5} \ , \\[2pt]
 \tilde{B}_{2,0}^{v,(0)}        &= 8 m_0 l_{1,1} \ , \\[2pt]
 \tilde{B}_{2,0}^{v,(m2)}(\mu)  &= -\frac{g_A^2 m_0 l_{1,1}}{2 F_\pi^2 \pi^2}+32 m_0 l^{(r)}_{3,1}(\mu)-32 l_{1,1} c_1 \ ,  \raisetag{18pt} \\[2pt]
 \tilde{B}_{2,0}^{v,(t)}        &= -8 m_0 l_{3,2} \ .
\end{align}
\end{subequations}
\section{Full results by diagram \label{app_full_result}}
In the following we will present the full one-loop results for the generalized form factors. To do this in the most economic way, we introduce the dimensionless quantities,
\begin{subequations}
\begin{align}
 \alpha&=\frac{m_\pi}{m_N} \ ,\\
 \tau&=\frac{t}{4 m_N^2} \ ,\\
 \tilde{\tau}&=\tau(1-4 u^2) \ .
\end{align}
\end{subequations}
We split the result for each form factor $X$ in parts orginating from the different Feynman diagrams shown in Fig.~\ref{FD_nGPDs},
\begin{align}
 X^{s/v} &= X^{s/v}_a + X^{s/v}_b + X^{s/v}_{de} + \int \limits_{-\frac{1}{2}}^{\frac{1}{2}} \!\! du \ \bigl( X^{s/v}_c+X^{s/v}_f \bigr) \ ,
\end{align}
where the triangle diagrams still have to be integrated over the parameter $u$. For diagrams $(c)$, $(de)$ and $(f)$ we sort the contributions by LECs and the different occurring structures. Schematically,
\begin{subequations}
 \begin{align}
 X^{v}_b &= \text{Prefactor} \times \text{LEC} \times  f^{v}_{b} \ , \\
 X^{s/v}_c &= \text{Prefactor} \times \sum_k \ \text{LEC}_k \ m_N^{n_k} \times  f^{X,s/v}_{ck} \ , \\
 X^{s/v}_{de} &= \text{Prefactor} \times \sum_k \ \text{LEC}_k \ m_N^{n_k} \times  f^{X,s/v}_{dek} \ , \\
 X^{s}_{f} &= \text{Prefactor} \times \text{LEC} \times f^{X,s}_{f} \ ,
\end{align}
\end{subequations}
where $n_k$ is chosen in such a way that the combination $\text{LEC}_k \ m_N^{n_k}$ is dimensionless. Note that the LEC occurring in diagrams of type  \subref{fd:subfigf} is no free fit parameter. It is connected to the chiral limit of the generalized form factors of the pion given in Eq.~\eqref{result_GFF_pion} by $l^s=A^{\pi,s,(0)}_{2,0}F_\pi^2/8$. The prefactors are tuned in order to have dimensionless  functions $f^{X,s/v}_{ck}$, $f^{X,s/v}_{dek}$ and $f^{X,s}_{f}$, given by
\begin{widetext}
\begin{subequations}
\begin{align}
f^v_b&=32 \pi^2 L+\ln \left(\frac{m_\pi^2}{\mu^2}\right) \ , \\
f^{X,s/v}_{ck} &= f^{X,s/v}_{ck1} + f^{X,s/v}_{ck2} 16 \pi^2 L + f^{X,s/v}_{ck3} \sqrt{4-\alpha ^2} \arccos\left(-\frac{\alpha }{2}\right) + f^{X,s/v}_{ck4} \frac{\arccos\left(-\frac{\alpha }{2 \sqrt{1-\tilde{\tau}}}\right)}{\sqrt{-\alpha ^2-4 \tilde{\tau}+4}} +f^{X,s/v}_{ck5} \ln \left(\frac{m_\pi^2}{\mu^2}\right)\ , \\
f^{X,s/v}_{dek} &= f^{X,s/v}_{dek1} + f^{X,s/v}_{dek2} 16 \pi^2 L + f^{X,s/v}_{dek3} \sqrt{4-\alpha ^2} \arccos\left(-\frac{\alpha }{2}\right) + f^{X,s/v}_{dek4} \ln \left(\frac{m_\pi^2}{\mu^2}\right)\ , \\
f^{X,s}_{f} &= f^{X,s}_{f1} + f^{X,s}_{f2} 16 \pi^2 L + f^{X,s}_{f3} \frac{\arccos\left(\frac{2 \tilde{\tau}-\alpha ^2}{2 \sqrt{1-\tilde{\tau}} \sqrt{\alpha ^2-\tilde{\tau}}}\right)}{\sqrt{-\alpha ^4+4 \alpha ^2-4 \tilde{\tau}}} + f^{X,s}_{f4} \ln \left(\frac{m_N^2 \left(\alpha ^2-\tilde{\tau}\right)}{\mu^2}\right)\ ,
\end{align}
\end{subequations}
where the $f^{X,s/v}_{cki}$, $f^{X,s/v}_{deki}$ and $f^{X,s/v}_{fi}$ are polynomials of $\alpha$, $\tau$, $\tilde{\tau}$ and $u$. The $Z$ factor for the nucleon is given by
\begin{align}
Z &= 1 + \frac{3 g_A^2 m_N^2}{32 \pi^2 F_\pi^2}\left(-\alpha ^2-\frac{2 \left(\alpha ^2-3\right) \alpha ^3 \arccos\left(-\frac{\alpha }{2}\right)}{\sqrt{4-\alpha ^2}}+16 \pi^2 \left(2 \alpha ^4-3 \alpha ^2\right) L+\frac{1}{2} \left(2 \alpha ^2-3\right) \alpha ^2 \ln \left(\frac{m_\pi^2}{\mu^2}\right)\right) \ .
\end{align}
\end{widetext}
In the results presented below we have appended the $Z$ factor to the leading and next-to-leading tree-level contributions only. However, not truncating the results, one could equitably argue that the $Z$ factor has to be appended as an overall prefactor to all diagrams.

\subsection{\texorpdfstring{$A^s_{20}$}{A20s}}
\begin{subequations}
\begin{align}
A^s_{a} &= 32 \alpha ^2 l^{s}_{2,2} m_N^2-32 \tau  l^{s}_{2,3} m_N^2+8 Z l^{s}_{0,1} \ , \\
A^s_{c} &= \frac{g_A^2 m_N^2}{16 \pi^2 F_\pi^2 (1-\tilde{\tau})^4}\bigl(f^{A,s}_{c1} l^{s}_{0,1}+f^{A,s}_{c2} l^{s}_{1,2} m_N\bigr) \ , \\
A^s_{de} &= \frac{g_A m_N^2}{16 \pi^2 F_\pi^2}\bigl(
\begin{aligned}[t]&
f^{A,s}_{de1} l^{s}_{1,6}+f^{A,s}_{de2} l^{s}_{1,13} m_N^2+f^{A,s}_{de3} l^{s}_{1,15} m_N^2\\
&+f^{A,s}_{de4} l^{s}_{1,18} m_N\bigr) \ ,
\end{aligned}
\raisetag{10pt}\\
 A^s_{f} &= \frac{g_A^2 l^{s} m_N^2}{16 \pi^2 F_\pi^4 (1-\tilde{\tau})^4}f^{A,s}_{f} \ ,
\end{align}
\end{subequations}
\begin{align*}
f^{A,s}_{c11} &= \frac{2}{3} \alpha ^6 \left(7 \tilde{\tau}^4-28 \tilde{\tau}^3+42 \tilde{\tau}^2-42 \tilde{\tau}-9\right)\\
&\quad-2 \alpha ^4 (\tilde{\tau}-1) \tilde{\tau} \left(7 \tilde{\tau}^2-21 \tilde{\tau}+32\right)-12 \alpha ^2 (\tilde{\tau}-1)^3\ , \\
f^{A,s}_{c12} &= -8 \alpha ^6 \left(\tilde{\tau}^4-4 \tilde{\tau}^3+6 \tilde{\tau}^2-6 \tilde{\tau}-3\right)\\
&\quad+36 \alpha ^4 (\tilde{\tau}-1) \left(\tilde{\tau}^3-3 \tilde{\tau}^2+4 \tilde{\tau}+2\right)\\
&\quad-12 \alpha ^2 (\tilde{\tau}-3) (\tilde{\tau}-1)^2 (\tilde{\tau}+1)\ , \\
f^{A,s}_{c13} &= 20 \alpha ^3 (\tilde{\tau}-1)^4-8 \alpha ^5 (\tilde{\tau}-1)^4\ , \\
f^{A,s}_{c14} &= -16 \alpha ^7 (\tilde{\tau}+2)-4 \alpha ^5 (\tilde{\tau}-1) (17 \tilde{\tau}+43)\\
&\quad-40 \alpha ^3 (\tilde{\tau}-1)^2 (\tilde{\tau}+5)\ , \\
f^{A,s}_{c15} &= -4 \alpha ^6 \left(\tilde{\tau}^4-4 \tilde{\tau}^3+6 \tilde{\tau}^2-6 \tilde{\tau}-3\right)\\
&\quad+18 \alpha ^4 (\tilde{\tau}-1) \left(\tilde{\tau}^3-3 \tilde{\tau}^2+4 \tilde{\tau}+2\right)\\
&\quad-6 \alpha ^2 (\tilde{\tau}-3) (\tilde{\tau}-1)^2 (\tilde{\tau}+1)\ , \\
f^{A,s}_{c21} &= -40 \alpha ^6 \tau -72 \alpha ^4 \tau  (\tilde{\tau}-1)\ , \\
f^{A,s}_{c22} &= 96 \alpha ^6 \tau +288 \alpha ^4 \tau  (\tilde{\tau}-1)+96 \alpha ^2 \tau  (\tilde{\tau}-1)^2\ , \\
f^{A,s}_{c23} &= 0\ , \\
f^{A,s}_{c24} &= -96 \alpha ^7 \tau -480 \alpha ^5 \tau  (\tilde{\tau}-1)-480 \alpha ^3 \tau  (\tilde{\tau}-1)^2\ , \\
f^{A,s}_{c25} &= 48 \alpha ^6 \tau +144 \alpha ^4 \tau  (\tilde{\tau}-1)+48 \alpha ^2 \tau  (\tilde{\tau}-1)^2\ , \\
 f^{A,s}_{de11} &= 16 \alpha ^4-\frac{28 \alpha ^6}{3}\ , \\
 f^{A,s}_{de12} &= 16 \alpha ^6-48 \alpha ^4\ , \\
 f^{A,s}_{de13} &= 16 \alpha ^5-16 \alpha ^3\ , \\
 f^{A,s}_{de14} &= 8 \alpha ^6-24 \alpha ^4\ , \\
 f^{A,s}_{de21} &= 16 \alpha ^8 \tau -\frac{256 \alpha ^6 \tau }{3}+2 \alpha ^4 (38 \tau -3)\ , \\
 f^{A,s}_{de22} &= -24 \alpha ^8 \tau +160 \alpha ^6 \tau -120 \alpha ^4 (2 \tau -1)\ , \\
 f^{A,s}_{de23} &= -24 \alpha ^7 \tau +112 \alpha ^5 \tau -64 \alpha ^3 \tau\ , \\
 f^{A,s}_{de24} &= -12 \alpha ^8 \tau +80 \alpha ^6 \tau -60 \alpha ^4 (2 \tau -1)\ , \\
 f^{A,s}_{de31} &= \frac{28 \alpha ^8}{3}-64 \alpha ^6+84 \alpha ^4\ , \\
 f^{A,s}_{de32} &= -16 \alpha ^8+144 \alpha ^6-336 \alpha ^4\ , \\
 f^{A,s}_{de33} &= -16 \alpha ^7+112 \alpha ^5-192 \alpha ^3\ , \\
 f^{A,s}_{de34} &= -8 \alpha ^8+72 \alpha ^6-168 \alpha ^4\ , \\
 f^{A,s}_{de41} &= \frac{28 \alpha ^8}{3}-64 \alpha ^6+96 \alpha ^4\ , \\
 f^{A,s}_{de42} &= -16 \alpha ^8+144 \alpha ^6-384 \alpha ^4\ , \\
 f^{A,s}_{de43} &= -16 \alpha ^7+112 \alpha ^5-192 \alpha ^3\ , \\
 f^{A,s}_{de44} &= -8 \alpha ^8+72 \alpha ^6-192 \alpha ^4\ , \\
 f^{A,s}_{f1} &= \frac{4}{3} \alpha ^6 (7 \tilde{\tau}+8)-4 \alpha ^4 \left(7 \tilde{\tau}^2+21 \tilde{\tau}+2\right)\\
&\quad+8 \alpha ^2 \tilde{\tau} \left(2 \tilde{\tau}^2+25 \tilde{\tau}+3\right)+\frac{16}{3} \tilde{\tau}^2 \left(\tilde{\tau}^2-28 \tilde{\tau}-3\right)\ , 
\\
 f^{A,s}_{f2} &= -16 \alpha ^6 (\tilde{\tau}+2)+24 \alpha ^4 \left(\tilde{\tau}^2+8 \tilde{\tau}+3\right)\\
&\quad-288 \alpha ^2 \tilde{\tau} (\tilde{\tau}+1)-16 \tilde{\tau}^2 \left(\tilde{\tau}^2-10 \tilde{\tau}-15\right)\ , 
\\
 f^{A,s}_{f3} &= 16 \alpha ^8 (\tilde{\tau}+2)-8 \alpha ^6 \left(3 \tilde{\tau}^2+28 \tilde{\tau}+17\right)\\
&\quad+16 \alpha ^4 \left(23 \tilde{\tau}^2+44 \tilde{\tau}+5\right)-16 \alpha ^2 \tilde{\tau} \left(9 \tilde{\tau}^2+74 \tilde{\tau}+13\right)\\
&\quad+128 \tilde{\tau}^2 (5 \tilde{\tau}+1)\ , 
\\
 f^{A,s}_{f4} &= -8 \alpha ^6 (\tilde{\tau}+2)+12 \alpha ^4 \left(\tilde{\tau}^2+8 \tilde{\tau}+3\right)\\
&\quad-144 \alpha ^2 \tilde{\tau} (\tilde{\tau}+1)-8 \tilde{\tau}^2 \left(\tilde{\tau}^2-10 \tilde{\tau}-15\right) \ .
\end{align*}
\subsection{\texorpdfstring{$B^s_{20}$}{B20s}}
\begin{subequations}
\begin{align}
B^s_{a} &= 64 \alpha ^2 l^{s}_{3,3} m_N^3-64 \tau  l^{s}_{3,4} m_N^3+16 Z l^{s}_{1,2} m_N \ ,\\
B^s_{c} &= \frac{g_A^2 m_N^2}{16 \pi^2 F_\pi^2 (1-\tilde{\tau})^4}\bigl(f^{B,s}_{c1} l^{s}_{0,1}+f^{B,s}_{c2} l^{s}_{1,2} m_N\bigr) \ ,\\
B^s_{de} &= \frac{g_A m_N^2}{16 \pi^2 F_\pi^2}\bigl(
\begin{aligned}[t]
&f^{B,s}_{de1} l^{s}_{1,13} m_N^2+f^{B,s}_{de2} l^{s}_{1,15} m_N^2\\
&+f^{B,s}_{de3} l^{s}_{1,18} m_N\bigr) \ ,
\end{aligned}
\\
 B^s_{f} &= \frac{g_A^2 l^{s} m_N^2}{16 \pi^2 F_\pi^4 (1-\tilde{\tau})^4}f^{B,s}_{f} \ ,
\end{align}
\end{subequations}
\begin{align*}
f^{B,s}_{c11} &= 20 \alpha ^6+36 \alpha ^4 (\tilde{\tau}-1)\ , \\
f^{B,s}_{c12} &= -48 \alpha ^6-144 \alpha ^4 (\tilde{\tau}-1)-48 \alpha ^2 (\tilde{\tau}-1)^2\ , \\
f^{B,s}_{c13} &= 0\ , \\
f^{B,s}_{c14} &= 48 \alpha ^7+240 \alpha ^5 (\tilde{\tau}-1)+240 \alpha ^3 (\tilde{\tau}-1)^2\ , \\
f^{B,s}_{c15} &= -24 \alpha ^6-72 \alpha ^4 (\tilde{\tau}-1)-24 \alpha ^2 (\tilde{\tau}-1)^2\ , \\
f^{B,s}_{c21} &= -\frac{4}{3} \alpha ^6 \left(7 \tilde{\tau}^4-28 \tilde{\tau}^3+42 \tilde{\tau}^2-37 \tilde{\tau}-14\right)\\
&\quad+4 \alpha ^4 (\tilde{\tau}-1) \left(7 \tilde{\tau}^3-21 \tilde{\tau}^2+21 \tilde{\tau}+11\right)\\
&\quad-24 \alpha ^2 (\tilde{\tau}-1)^3\ , \\
f^{B,s}_{c22} &= 16 \alpha ^6 \left(\tilde{\tau}^4-4 \tilde{\tau}^3+6 \tilde{\tau}^2-7 \tilde{\tau}-2\right)\\
&\quad-72 \alpha ^4 (\tilde{\tau}-1) \left(\tilde{\tau}^3-3 \tilde{\tau}^2+5 \tilde{\tau}+1\right)\\
&\quad+24 \alpha ^2 (\tilde{\tau}-1)^2 \left(\tilde{\tau}^2-4 \tilde{\tau}-1\right)\ , \\
f^{B,s}_{c23} &= 16 \alpha ^5 (\tilde{\tau}-1)^4-40 \alpha ^3 (\tilde{\tau}-1)^4\ , \\
f^{B,s}_{c24} &= 48 \alpha ^7 (\tilde{\tau}+1)+240 \alpha ^5 (\tilde{\tau}-1) (\tilde{\tau}+1)\\
&\quad+240 \alpha ^3 (\tilde{\tau}-1)^2 (\tilde{\tau}+1)\ , \\
f^{B,s}_{c25} &= 8 \alpha ^6 \left(\tilde{\tau}^4-4 \tilde{\tau}^3+6 \tilde{\tau}^2-7 \tilde{\tau}-2\right)\\
&\quad-36 \alpha ^4 (\tilde{\tau}-1) \left(\tilde{\tau}^3-3 \tilde{\tau}^2+5 \tilde{\tau}+1\right)\\
&\quad+12 \alpha ^2 (\tilde{\tau}-1)^2 \left(\tilde{\tau}^2-4 \tilde{\tau}-1\right)\ , \\
 f^{B,s}_{de11} &= -\frac{10 \alpha ^8}{3}+\frac{88 \alpha ^6}{3}-70 \alpha ^4\ , \\
 f^{B,s}_{de12} &= 4 \alpha ^8-40 \alpha ^6+120 \alpha ^4\ , \\
 f^{B,s}_{de13} &= 4 \alpha ^7-32 \alpha ^5+64 \alpha ^3\ , \\
 f^{B,s}_{de14} &= 2 \alpha ^8-20 \alpha ^6+60 \alpha ^4\ , \\
 f^{B,s}_{de21} &= -\frac{28 \alpha ^8}{3}+64 \alpha ^6-96 \alpha ^4\ , \\
 f^{B,s}_{de22} &= 16 \alpha ^8-144 \alpha ^6+384 \alpha ^4\ , \\
 f^{B,s}_{de23} &= 16 \alpha ^7-112 \alpha ^5+192 \alpha ^3\ , \\
 f^{B,s}_{de24} &= 8 \alpha ^8-72 \alpha ^6+192 \alpha ^4\ , \\
 f^{B,s}_{de31} &= -16 \alpha ^8+\frac{256 \alpha ^6}{3}-76 \alpha ^4\ , \\
 f^{B,s}_{de32} &= 24 \alpha ^8-160 \alpha ^6+240 \alpha ^4\ , \\
 f^{B,s}_{de33} &= 24 \alpha ^7-112 \alpha ^5+64 \alpha ^3\ , \\
 f^{B,s}_{de34} &= 12 \alpha ^8-80 \alpha ^6+120 \alpha ^4\ , \\
 f^{B,s}_{f1} &= -20 \alpha ^6+12 \alpha ^4 (7 \tilde{\tau}+3)-120 \alpha ^2 \tilde{\tau} (\tilde{\tau}+1)\\
&\quad+32 \tilde{\tau}^2 (2 \tilde{\tau}+3)\ , 
\\
 f^{B,s}_{f2} &= 48 \alpha ^6-144 \alpha ^4 (\tilde{\tau}+1)+48 \alpha ^2 \left(3 \tilde{\tau}^2+8 \tilde{\tau}+1\right)\\
&\quad-48 \tilde{\tau} \left(\tilde{\tau}^2+6 \tilde{\tau}+1\right)\ , 
\\
 f^{B,s}_{f3} &= -48 \alpha ^8+48 \alpha ^6 (3 \tilde{\tau}+5)-48 \alpha ^4 (\tilde{\tau}+5) (3 \tilde{\tau}+1)\\
&\quad+48 \alpha ^2 \tilde{\tau} \left(\tilde{\tau}^2+18 \tilde{\tau}+13\right)-384 \tilde{\tau}^2 (\tilde{\tau}+1)\ , 
\\
 f^{B,s}_{f4} &= 24 \alpha ^6-72 \alpha ^4 (\tilde{\tau}+1)+24 \alpha ^2 \left(3 \tilde{\tau}^2+8 \tilde{\tau}+1\right)\\
&\quad-24 \tilde{\tau} \left(\tilde{\tau}^2+6 \tilde{\tau}+1\right)\ . 
\end{align*}
\subsection{\texorpdfstring{$C^s_{2}$}{C2s}}
\begin{subequations}
\begin{align}
C^s_{a} &= -4 l^{s}_{2,1} m_N  \ , \\
C^s_{c} &= \frac{g_A^2 m_N^2}{16 \pi^2 F_\pi^2 (1-\tilde{\tau})^4}\bigl(f^{C,s}_{c1} l^{s}_{0,1}+f^{C,s}_{c2} l^{s}_{1,2} m_N\bigr) \ , \\
C^s_{de} &= \frac{g_A m_N^2}{16 \pi^2 F_\pi^2}\bigl(
\begin{aligned}[t]
 & f^{C,s}_{de1} l^{s}_{1,13} m_N^2+f^{C,s}_{de2} l^{s}_{1,15} m_N^2\\
&+f^{C,s}_{de3} l^{s}_{1,18} m_N\bigr)  \ ,
\end{aligned} \\
 C^s_{f} &= \frac{g_A^2 l^{s} m_N^2}{16 \pi^2 F_\pi^4 (1-\tilde{\tau})^4}f^{C,s}_{f}  \ ,
\end{align}
\end{subequations}
\begin{align*}
f^{C,s}_{c11} &= -20 \alpha ^6 u^2-24 \alpha ^4 (\tilde{\tau}-1) u^2\ , \\
f^{C,s}_{c12} &= 48 \alpha ^6 u^2+96 \alpha ^4 (\tilde{\tau}-1) u^2\ , \\
f^{C,s}_{c13} &= 0\ , \\
f^{C,s}_{c14} &= -48 \alpha ^7 u^2-192 \alpha ^5 (\tilde{\tau}-1) u^2-96 \alpha ^3 (\tilde{\tau}-1)^2 u^2\ , \\
f^{C,s}_{c15} &= 24 \alpha ^6 u^2+48 \alpha ^4 (\tilde{\tau}-1) u^2\ , \\
f^{C,s}_{c21} &= \frac{7}{3} \alpha ^6 (\tilde{\tau}-1)+4 \alpha ^4 (\tilde{\tau}-1)^2\ , \\
f^{C,s}_{c22} &= -4 \alpha ^6 (\tilde{\tau}-1)-12 \alpha ^4 (\tilde{\tau}-1)^2\ , \\
f^{C,s}_{c23} &= 0\ , \\
f^{C,s}_{c24} &= 4 \alpha ^7 (\tilde{\tau}-1)+20 \alpha ^5 (\tilde{\tau}-1)^2+16 \alpha ^3 (\tilde{\tau}-1)^3\ , \\
f^{C,s}_{c25} &= -2 \alpha ^6 (\tilde{\tau}-1)-6 \alpha ^4 (\tilde{\tau}-1)^2\ , \\
 f^{C,s}_{de11} &= -\frac{5 \alpha ^8}{6}+4 \alpha ^6-\frac{3 \alpha ^4}{2}\ , \\
 f^{C,s}_{de12} &= \alpha ^8-6 \alpha ^6+6 \alpha ^4\ , \\
 f^{C,s}_{de13} &= \alpha ^7-4 \alpha ^5\ , \\
 f^{C,s}_{de14} &= \frac{\alpha ^8}{2}-3 \alpha ^6+3 \alpha ^4\ , \\
 f^{C,s}_{de21} &= \frac{7 \alpha ^8}{3}-4 \alpha ^6\ , \\
 f^{C,s}_{de22} &= 12 \alpha ^6-4 \alpha ^8\ , \\
 f^{C,s}_{de23} &= 4 \alpha ^5-4 \alpha ^7\ , \\
 f^{C,s}_{de24} &= 6 \alpha ^6-2 \alpha ^8\ , \\
 f^{C,s}_{de31} &= 4 \alpha ^8-12 \alpha ^6+3 \alpha ^4\ , \\
 f^{C,s}_{de32} &= -6 \alpha ^8+24 \alpha ^6-12 \alpha ^4\ , \\
 f^{C,s}_{de33} &= 12 \alpha ^5-6 \alpha ^7\ , \\
 f^{C,s}_{de34} &= -3 \alpha ^8+12 \alpha ^6-6 \alpha ^4\ , \\
 f^{C,s}_{f1} &= 6 (\tilde{\tau}-1)^3 \tilde{\tau}+20 \alpha ^6 u^2-60 \alpha ^4 (\tilde{\tau}+1) u^2\\
&\quad+24 \alpha ^2 \tilde{\tau} (\tilde{\tau}+9) u^2-8 \tilde{\tau} \left(3 \tilde{\tau}^3-13 \tilde{\tau}^2+33 \tilde{\tau}-3\right) u^2\ , 
\\
 f^{C,s}_{f2} &= 12 (\tilde{\tau}-3) (\tilde{\tau}-1)^2 \tilde{\tau}-48 \alpha ^6 u^2+48 \alpha ^4 (\tilde{\tau}+5) u^2\\
&\quad+\alpha ^2 \left(12 (\tilde{\tau}-1)^2-288 (\tilde{\tau}+1) u^2\right)\\
&\quad-48 \tilde{\tau} \left(\tilde{\tau}^3-4 \tilde{\tau}^2+5 \tilde{\tau}-10\right) u^2\ , 
\\
 f^{C,s}_{f3} &= -48 (\tilde{\tau}-1)^2 \tilde{\tau}+48 \alpha ^8 u^2-48 \alpha ^6 (\tilde{\tau}+7) u^2\\
&\quad+\alpha ^4 \left(96 (5 \tilde{\tau}+7) u^2-12 (\tilde{\tau}-1)^2\right)\\
&\quad+\alpha ^2 \left(12 (\tilde{\tau}-1)^2 (\tilde{\tau}+3)-48 (\tilde{\tau}+7) (3 \tilde{\tau}+1) u^2\right)\\
&\quad+384 \tilde{\tau} (\tilde{\tau}+1) u^2\ , 
\\
 f^{C,s}_{f4} &= 6 (\tilde{\tau}-3) (\tilde{\tau}-1)^2 \tilde{\tau}-24 \alpha ^6 u^2+24 \alpha ^4 (\tilde{\tau}+5) u^2\\
&\quad+\alpha ^2 \left(6 (\tilde{\tau}-1)^2-144 (\tilde{\tau}+1) u^2\right)\\
&\quad-24 \tilde{\tau} \left(\tilde{\tau}^3-4 \tilde{\tau}^2+5 \tilde{\tau}-10\right) u^2 \ .
\end{align*}
\subsection{\texorpdfstring{$\tilde{A}^s_{20}$}{At20s}}
\begin{subequations}
\begin{align}
\tilde{A}^s_{a} &= 32 \alpha ^2 l^{s}_{2,4} m_N^2-32 \tau  l^{s}_{2,5} m_N^2+8 Z l^{s}_{0,2}  \ , \\
\tilde{A}^s_{c} &= \frac{g_A^2 m_N^2}{16 \pi^2 F_\pi^2 (1-\tilde{\tau})^4} f^{\tilde{A},s}_{c1} l^{s}_{0,2}  \ , \\
\tilde{A}^s_{de} &= \frac{g_A m_N^2}{16 \pi^2 F_\pi^2}\big(f^{\tilde{A},s}_{de1} l^{s}_{1,4}+f^{\tilde{A},s}_{de2} l^{s}_{1,9} m_N^2+f^{\tilde{A},s}_{de3} l^{s}_{1,11} m_N^2\big)  \ ,
\end{align}
\end{subequations}
\begin{align*}
f^{\tilde{A},s}_{c11} &= \frac{14}{3} \alpha ^6 (\tilde{\tau}-1) \left(\tilde{\tau}^3-3 \tilde{\tau}^2+3 \tilde{\tau}+1\right)\\
&\quad-2 \alpha ^4 (\tilde{\tau}-1)^2 \left(7 \tilde{\tau}^2-14 \tilde{\tau}-4\right)+12 \alpha ^2 (\tilde{\tau}-1)^3\ , \\
f^{\tilde{A},s}_{c12} &= -8 \alpha ^6 (\tilde{\tau}-1) \left(\tilde{\tau}^3-3 \tilde{\tau}^2+3 \tilde{\tau}+1\right)\\
&\quad+36 \alpha ^4 (\tilde{\tau}-2) (\tilde{\tau}-1)^2 \tilde{\tau}-12 \alpha ^2 (\tilde{\tau}-1)^4\ , \\
f^{\tilde{A},s}_{c13} &= 20 \alpha ^3 (\tilde{\tau}-1)^4-8 \alpha ^5 (\tilde{\tau}-1)^4\ , \\
f^{\tilde{A},s}_{c14} &= 16 \alpha ^7 (\tilde{\tau}-1)+68 \alpha ^5 (\tilde{\tau}-1)^2+40 \alpha ^3 (\tilde{\tau}-1)^3\ , \\
f^{\tilde{A},s}_{c15} &= -4 \alpha ^6 (\tilde{\tau}-1) \left(\tilde{\tau}^3-3 \tilde{\tau}^2+3 \tilde{\tau}+1\right)\\
&\quad+18 \alpha ^4 (\tilde{\tau}-2) (\tilde{\tau}-1)^2 \tilde{\tau}-6 \alpha ^2 (\tilde{\tau}-1)^4\ , \\
 f^{\tilde{A},s}_{de11} &= 16 \alpha ^4-\frac{28 \alpha ^6}{3}\ , \\
 f^{\tilde{A},s}_{de12} &= 16 \alpha ^6-48 \alpha ^4\ , \\
 f^{\tilde{A},s}_{de13} &= 16 \alpha ^5-16 \alpha ^3\ , \\
 f^{\tilde{A},s}_{de14} &= 8 \alpha ^6-24 \alpha ^4\ , \\
 f^{\tilde{A},s}_{de21} &= 16 \alpha ^8 \tau -\frac{256 \alpha ^6 \tau }{3}+2 \alpha ^4 (38 \tau -3)\ , \\
 f^{\tilde{A},s}_{de22} &= -24 \alpha ^8 \tau +160 \alpha ^6 \tau -120 \alpha ^4 (2 \tau -1)\ , \\
 f^{\tilde{A},s}_{de23} &= -24 \alpha ^7 \tau +112 \alpha ^5 \tau -64 \alpha ^3 \tau\ , \\
 f^{\tilde{A},s}_{de24} &= -12 \alpha ^8 \tau +80 \alpha ^6 \tau -60 \alpha ^4 (2 \tau -1)\ , \\
 f^{\tilde{A},s}_{de31} &= -12 \alpha ^4\ , \\
 f^{\tilde{A},s}_{de32} &= 48 \alpha ^4\ , \\
 f^{\tilde{A},s}_{de33} &= 0\ , \\
 f^{\tilde{A},s}_{de34} &= 24 \alpha ^4\ .
\end{align*}
\subsection{\texorpdfstring{$\tilde{B}^s_{20}$}{Bt20s}}
\begin{subequations}
\begin{align}
\tilde{B}^s_{a} &= 64 \alpha ^2 l^{s}_{3,1} m_N^3-64 \tau  l^{s}_{3,2} m_N^3+16 Z l^{s}_{1,1} m_N  \ , \\
\tilde{B}^s_{c} &= \frac{g_A^2 m_N^2}{16 \pi^2 F_\pi^2 (1-\tilde{\tau})^4}\bigl(f^{\tilde{B},s}_{c1} l^{s}_{0,2}+f^{\tilde{B},s}_{c2} l^{s}_{1,1} m_N\bigr) \ , \\
\tilde{B}^s_{de} &= \frac{g_A m_N^2}{16 \pi^2 F_\pi^2}\bigl(f^{\tilde{B},s}_{de1} l^{s}_{1,9} m_N^2+f^{\tilde{B},s}_{de2} l^{s}_{1,11} m_N^2\bigr)  \ ,
\end{align}
\end{subequations}
\begin{align*}
f^{\tilde{B},s}_{c11} &= 160 \alpha ^6 u^2+240 \alpha ^4 (\tilde{\tau}-1) u^2\ , \\
f^{\tilde{B},s}_{c12} &= -384 \alpha ^6 u^2-960 \alpha ^4 (\tilde{\tau}-1) u^2-192 \alpha ^2 (\tilde{\tau}-1)^2 u^2\ , \\
f^{\tilde{B},s}_{c13} &= 0\ , \\
f^{\tilde{B},s}_{c14} &= 384 \alpha ^7 u^2+1728 \alpha ^5 (\tilde{\tau}-1) u^2+1344 \alpha ^3 (\tilde{\tau}-1)^2 u^2\ , \\
f^{\tilde{B},s}_{c15} &= -192 \alpha ^6 u^2-480 \alpha ^4 (\tilde{\tau}-1) u^2-96 \alpha ^2 (\tilde{\tau}-1)^2 u^2\ , \\
f^{\tilde{B},s}_{c21} &= -\frac{28}{3} \alpha ^6 (\tilde{\tau}-1)^4+28 \alpha ^4 (\tilde{\tau}-1)^4+24 \alpha ^2 (\tilde{\tau}-1)^3\ , \\
f^{\tilde{B},s}_{c22} &= 16 \alpha ^6 (\tilde{\tau}-1)^4-24 \alpha ^4 (\tilde{\tau}-1)^2 \left(3 \tilde{\tau}^2-6 \tilde{\tau}+1\right)\\
&\quad+24 \alpha ^2 (\tilde{\tau}-1)^3 (\tilde{\tau}+1)\ , \\
f^{\tilde{B},s}_{c23} &= 16 \alpha ^5 (\tilde{\tau}-1)^4-40 \alpha ^3 (\tilde{\tau}-1)^4\ , \\
f^{\tilde{B},s}_{c24} &= -48 \alpha ^5 (\tilde{\tau}-1)^2-144 \alpha ^3 (\tilde{\tau}-1)^3\ , \\
f^{\tilde{B},s}_{c25} &= 8 \alpha ^6 (\tilde{\tau}-1)^4-12 \alpha ^4 (\tilde{\tau}-1)^2 \left(3 \tilde{\tau}^2-6 \tilde{\tau}+1\right)\\
&\quad+12 \alpha ^2 (\tilde{\tau}-1)^3 (\tilde{\tau}+1)\ , \\
 f^{\tilde{B},s}_{de11} &= \frac{40 \alpha ^6}{3}-64 \alpha ^4\ , \\
 f^{\tilde{B},s}_{de12} &= 96 \alpha ^4-16 \alpha ^6\ , \\
 f^{\tilde{B},s}_{de13} &= 64 \alpha ^3-16 \alpha ^5\ , \\
 f^{\tilde{B},s}_{de14} &= 48 \alpha ^4-8 \alpha ^6\ , \\
 f^{\tilde{B},s}_{de21} &= \frac{56 \alpha ^8}{3}-80 \alpha ^6\ , \\
 f^{\tilde{B},s}_{de22} &= -32 \alpha ^8+192 \alpha ^6-192 \alpha ^4\ , \\
 f^{\tilde{B},s}_{de23} &= 128 \alpha ^5-32 \alpha ^7\ , \\
 f^{\tilde{B},s}_{de24} &= -16 \alpha ^8+96 \alpha ^6-96 \alpha ^4\ .
\end{align*}
\subsection{\texorpdfstring{$A^v_{20}$}{A20v}}
\begin{subequations}
\begin{align}
A^v_{a} &= 16 \alpha ^2 l_{2,2} m_N^2-16 \tau  l_{2,3} m_N^2+4 Z l_{0,1}  \ , \\
A^v_{b} &= -\frac{\alpha ^2 m_N^2 l_{0,1} }{4 \pi^2 F_\pi^2}f^v_b  \ , \\
A^v_{c} &= \frac{g_A^2 m_N^2}{16 \pi^2 F_\pi^2 (1-\tilde{\tau})^4}\bigl(f^{A,v}_{c1} l_{0,1}+f^{A,v}_{c2} l_{1,2} m_N\bigr)  \ , \\
A^v_{de} &= \frac{g_A m_N^2}{16 \pi^2 F_\pi^2}\bigl(
\begin{aligned}[t]
& f^{A,v}_{de1} l_{0,2}+f^{A,v}_{de2} l_{1,1} m_N+f^{A,v}_{de3} l_{1,3} m_N\\
&+f^{A,v}_{de4} l_{1,8} m_N^3+f^{A,v}_{de5} ( l_{1,6} +  l_{1,7} )\\
&+f^{A,v}_{de6} (l_{1,13}+l_{1,14}) m_N^2\\
&+f^{A,v}_{de7} (l_{1,15}+l_{1,16}) m_N^2\\
&+f^{A,v}_{de8} (l_{1,18}+l_{1,19}) m_N\bigr)  \ ,
\end{aligned} \raisetag{12pt}
\end{align}
\end{subequations}
\begin{align*}
f^{A,v}_{c11} &= \frac{1}{9} \alpha ^6 \left(-7 \tilde{\tau}^4+28 \tilde{\tau}^3-42 \tilde{\tau}^2+42 \tilde{\tau}+9\right)\\
&\quad+\frac{1}{3} \alpha ^4 (\tilde{\tau}-1) \tilde{\tau} \left(7 \tilde{\tau}^2-21 \tilde{\tau}+32\right)+2 \alpha ^2 (\tilde{\tau}-1)^3\ , \\
f^{A,v}_{c12} &= \frac{4}{3} \alpha ^6 \left(\tilde{\tau}^4-4 \tilde{\tau}^3+6 \tilde{\tau}^2-6 \tilde{\tau}-3\right)\\
&\quad-6 \alpha ^4 (\tilde{\tau}-1) \left(\tilde{\tau}^3-3 \tilde{\tau}^2+4 \tilde{\tau}+2\right)\\
&\quad+2 \alpha ^2 (\tilde{\tau}-3) (\tilde{\tau}-1)^2 (\tilde{\tau}+1)\ , \\
f^{A,v}_{c13} &= \frac{4}{3} \alpha ^5 (\tilde{\tau}-1)^4-\frac{10}{3} \alpha ^3 (\tilde{\tau}-1)^4\ , \\
f^{A,v}_{c14} &= \frac{8}{3} \alpha ^7 (\tilde{\tau}+2)+\frac{2}{3} \alpha ^5 (\tilde{\tau}-1) (17 \tilde{\tau}+43)\\
&\quad+\frac{20}{3} \alpha ^3 (\tilde{\tau}-1)^2 (\tilde{\tau}+5)\ , \\
f^{A,v}_{c15} &= \frac{2}{3} \alpha ^6 \left(\tilde{\tau}^4-4 \tilde{\tau}^3+6 \tilde{\tau}^2-6 \tilde{\tau}-3\right)\\
&\quad-3 \alpha ^4 (\tilde{\tau}-1) \left(\tilde{\tau}^3-3 \tilde{\tau}^2+4 \tilde{\tau}+2\right)\\
&\quad+\alpha ^2 (\tilde{\tau}-3) (\tilde{\tau}-1)^2 (\tilde{\tau}+1)\ , \\
f^{A,v}_{c21} &= \frac{20 \alpha ^6 \tau }{3}+12 \alpha ^4 \tau  (\tilde{\tau}-1)\ , \\
f^{A,v}_{c22} &= -16 \alpha ^6 \tau -48 \alpha ^4 \tau  (\tilde{\tau}-1)-16 \alpha ^2 \tau  (\tilde{\tau}-1)^2\ , \\
f^{A,v}_{c23} &= 0\ , \\
f^{A,v}_{c24} &= 16 \alpha ^7 \tau +80 \alpha ^5 \tau  (\tilde{\tau}-1)+80 \alpha ^3 \tau  (\tilde{\tau}-1)^2\ , \\
f^{A,v}_{c25} &= -8 \alpha ^6 \tau -24 \alpha ^4 \tau  (\tilde{\tau}-1)-8 \alpha ^2 \tau  (\tilde{\tau}-1)^2\ , \\
 f^{A,v}_{de11} &= \frac{7 \alpha ^6}{9}-\frac{10 \alpha ^4}{3}\ , \\
 f^{A,v}_{de12} &= 8 \alpha ^4-\frac{4 \alpha ^6}{3}\ , \\
 f^{A,v}_{de13} &= \frac{16 \alpha ^3}{3}-\frac{4 \alpha ^5}{3}\ , \\
 f^{A,v}_{de14} &= 4 \alpha ^4-\frac{2 \alpha ^6}{3}\ , \\
 f^{A,v}_{de21} &= \frac{7 \alpha ^8}{18}-\frac{5 \alpha ^6}{3}\ , \\
 f^{A,v}_{de22} &= -\frac{2 \alpha ^8}{3}+4 \alpha ^6-4 \alpha ^4\ , \\
 f^{A,v}_{de23} &= \frac{8 \alpha ^5}{3}-\frac{2 \alpha ^7}{3}\ , \\
 f^{A,v}_{de24} &= -\frac{\alpha ^8}{3}+2 \alpha ^6-2 \alpha ^4\ , \\
 f^{A,v}_{de31} &= \frac{10 \alpha ^6}{3}-\frac{7 \alpha ^8}{9}\ , \\
 f^{A,v}_{de32} &= \frac{4 \alpha ^8}{3}-8 \alpha ^6+8 \alpha ^4\ , \\
 f^{A,v}_{de33} &= \frac{4 \alpha ^7}{3}-\frac{16 \alpha ^5}{3}\ , \\
 f^{A,v}_{de34} &= \frac{2 \alpha ^8}{3}-4 \alpha ^6+4 \alpha ^4\ , \\
 f^{A,v}_{de41} &= \frac{2}{75} \alpha ^{10} (125 \tau -26)-\frac{8}{45} \alpha ^8 (135 \tau -22)\\
&\quad+\frac{2}{45} \alpha ^6 (1075 \tau -93)-\frac{76 \alpha ^4 \tau }{3}\ , \\
 f^{A,v}_{de42} &= -\frac{4}{5} \alpha ^{10} (5 \tau -1)+\frac{8}{3} \alpha ^8 (13 \tau -2)\\
&\quad-\frac{8}{3} \alpha ^6 (35 \tau -3)+16 \alpha ^4 (5 \tau -2)\ , \\
 f^{A,v}_{de43} &= -\frac{4}{5} \alpha ^9 (5 \tau -1)+\frac{8}{15} \alpha ^7 (50 \tau -7)\\
&\quad-\frac{16}{15} \alpha ^5 (45 \tau -2)+\frac{64 \alpha ^3 \tau }{3}\ , \\
 f^{A,v}_{de44} &= -\frac{2}{5} \alpha ^{10} (5 \tau -1)+\frac{4}{3} \alpha ^8 (13 \tau -2)\\
&\quad-\frac{4}{3} \alpha ^6 (35 \tau -3)+8 \alpha ^4 (5 \tau -2)\ , \\
 f^{A,v}_{de51} &= \frac{8 \alpha ^4}{3}-\frac{14 \alpha ^6}{9}\ , \\
 f^{A,v}_{de52} &= \frac{8 \alpha ^6}{3}-8 \alpha ^4\ , \\
 f^{A,v}_{de53} &= \frac{8 \alpha ^5}{3}-\frac{8 \alpha ^3}{3}\ , \\
 f^{A,v}_{de54} &= \frac{4 \alpha ^6}{3}-4 \alpha ^4\ , \\
 f^{A,v}_{de61} &= \frac{8 \alpha ^8 \tau }{3}-\frac{128 \alpha ^6 \tau }{9}+\frac{1}{3} \alpha ^4 (38 \tau -3)\ , \\
 f^{A,v}_{de62} &= -4 \alpha ^8 \tau +\frac{80 \alpha ^6 \tau }{3}-20 \alpha ^4 (2 \tau -1)\ , \\
 f^{A,v}_{de63} &= -4 \alpha ^7 \tau +\frac{56 \alpha ^5 \tau }{3}-\frac{32 \alpha ^3 \tau }{3}\ , \\
 f^{A,v}_{de64} &= -2 \alpha ^8 \tau +\frac{40 \alpha ^6 \tau }{3}-10 \alpha ^4 (2 \tau -1)\ , \\
 f^{A,v}_{de71} &= \frac{14 \alpha ^8}{9}-\frac{32 \alpha ^6}{3}+14 \alpha ^4\ , \\
 f^{A,v}_{de72} &= -\frac{8 \alpha ^8}{3}+24 \alpha ^6-56 \alpha ^4\ , \\
 f^{A,v}_{de73} &= -\frac{8 \alpha ^7}{3}+\frac{56 \alpha ^5}{3}-32 \alpha ^3\ , \\
 f^{A,v}_{de74} &= -\frac{4 \alpha ^8}{3}+12 \alpha ^6-28 \alpha ^4\ , \\
 f^{A,v}_{de81} &= \frac{14 \alpha ^8}{9}-\frac{32 \alpha ^6}{3}+16 \alpha ^4\ , \\
 f^{A,v}_{de82} &= -\frac{8 \alpha ^8}{3}+24 \alpha ^6-64 \alpha ^4\ , \\
 f^{A,v}_{de83} &= -\frac{8 \alpha ^7}{3}+\frac{56 \alpha ^5}{3}-32 \alpha ^3\ , \\
 f^{A,v}_{de84} &= -\frac{4 \alpha ^8}{3}+12 \alpha ^6-32 \alpha ^4\ .
\end{align*}
\subsection{\texorpdfstring{$B^v_{20}$}{B20v}}
\begin{subequations}
\begin{align}
B^v_{a} &= 32 \alpha ^2 l_{3,3} m_N^3-32 \tau  l_{3,4} m_N^3+8 Z l_{1,2} m_N  \ , \\
B^v_{b} &= -\frac{\alpha ^2 m_N^2 l_{1,2} m_N }{2 \pi^2 F_\pi^2}f^v_b \ , \\
B^v_{c} &= \frac{g_A^2 m_N^2}{16 \pi^2 F_\pi^2 (1-\tilde{\tau})^4}\bigl(f^{B,v}_{c1} l_{0,1}+f^{B,v}_{c2} l_{1,2} m_N\bigr) \ , \\
B^v_{de} &= \frac{g_A m_N^2}{16 \pi^2 F_\pi^2}\bigl(
\begin{aligned}[t]
&f^{B,v}_{de1} l_{1,1} m_N+f^{B,v}_{de2} l_{1,3} m_N+f^{B,v}_{de3} l_{1,8} m_N^3\\
&+f^{B,v}_{de4} (l_{1,13}+l_{1,14}) m_N^2\\
&+f^{B,v}_{de5} (l_{1,15}+l_{1,16}) m_N^2\\
&+f^{B,v}_{de6} (l_{1,18}+l_{1,19}) m_N\bigr) \ ,
\end{aligned} \raisetag{12pt}
\end{align}
\end{subequations}
\begin{align*}
f^{B,v}_{c11} &= -\frac{10 \alpha ^6}{3}-6 \alpha ^4 (\tilde{\tau}-1)\ , \\
f^{B,v}_{c12} &= 8 \alpha ^6+24 \alpha ^4 (\tilde{\tau}-1)+8 \alpha ^2 (\tilde{\tau}-1)^2\ , \\
f^{B,v}_{c13} &= 0\ , \\
f^{B,v}_{c14} &= -8 \alpha ^7-40 \alpha ^5 (\tilde{\tau}-1)-40 \alpha ^3 (\tilde{\tau}-1)^2\ , \\
f^{B,v}_{c15} &= 4 \alpha ^6+12 \alpha ^4 (\tilde{\tau}-1)+4 \alpha ^2 (\tilde{\tau}-1)^2\ , \\
f^{B,v}_{c21} &= \frac{2}{9} \alpha ^6 \left(7 \tilde{\tau}^4-28 \tilde{\tau}^3+42 \tilde{\tau}^2-37 \tilde{\tau}-14\right)\\
&\quad-\frac{2}{3} \alpha ^4 (\tilde{\tau}-1) \left(7 \tilde{\tau}^3-21 \tilde{\tau}^2+21 \tilde{\tau}+11\right)\\
&\quad+4 \alpha ^2 (\tilde{\tau}-1)^3\ , \\
f^{B,v}_{c22} &= -\frac{8}{3} \alpha ^6 \left(\tilde{\tau}^4-4 \tilde{\tau}^3+6 \tilde{\tau}^2-7 \tilde{\tau}-2\right)\\
&\quad+12 \alpha ^4 (\tilde{\tau}-1) \left(\tilde{\tau}^3-3 \tilde{\tau}^2+5 \tilde{\tau}+1\right)\\
&\quad-4 \alpha ^2 (\tilde{\tau}-1)^2 \left(\tilde{\tau}^2-4 \tilde{\tau}-1\right)\ , \\
f^{B,v}_{c23} &= \frac{20}{3} \alpha ^3 (\tilde{\tau}-1)^4-\frac{8}{3} \alpha ^5 (\tilde{\tau}-1)^4\ , \\
f^{B,v}_{c24} &= -8 \alpha ^7 (\tilde{\tau}+1)-40 \alpha ^5 (\tilde{\tau}-1) (\tilde{\tau}+1)\\
&\quad-40 \alpha ^3 (\tilde{\tau}-1)^2 (\tilde{\tau}+1)\ , \\
f^{B,v}_{c25} &= -\frac{4}{3} \alpha ^6 \left(\tilde{\tau}^4-4 \tilde{\tau}^3+6 \tilde{\tau}^2-7 \tilde{\tau}-2\right)\\
&\quad+6 \alpha ^4 (\tilde{\tau}-1) \left(\tilde{\tau}^3-3 \tilde{\tau}^2+5 \tilde{\tau}+1\right)\\
&\quad-2 \alpha ^2 (\tilde{\tau}-1)^2 \left(\tilde{\tau}^2-4 \tilde{\tau}-1\right)\ , \\
 f^{B,v}_{de11} &= \frac{2 \alpha ^8}{3}-\frac{4 \alpha ^6}{9}-\frac{13 \alpha ^4}{6}\ , \\
 f^{B,v}_{de12} &= -\alpha ^8+\frac{4 \alpha ^6}{3}+6 \alpha ^4\ , \\
 f^{B,v}_{de13} &= -\alpha ^7-\frac{2 \alpha ^5}{3}+\frac{8 \alpha ^3}{3}\ , \\
 f^{B,v}_{de14} &= -\frac{\alpha ^8}{2}+\frac{2 \alpha ^6}{3}+3 \alpha ^4\ , \\
 f^{B,v}_{de21} &= -\frac{4 \alpha ^8}{3}+\frac{8 \alpha ^6}{9}+\frac{13 \alpha ^4}{3}\ , \\
 f^{B,v}_{de22} &= 2 \alpha ^8-\frac{8 \alpha ^6}{3}-12 \alpha ^4\ , \\
 f^{B,v}_{de23} &= 2 \alpha ^7+\frac{4 \alpha ^5}{3}-\frac{16 \alpha ^3}{3}\ , \\
 f^{B,v}_{de24} &= \alpha ^8-\frac{4 \alpha ^6}{3}-6 \alpha ^4\ , \\
 f^{B,v}_{de31} &= -\frac{2}{3} \alpha ^{10} (7 \tau -2)+\frac{4}{27} \alpha ^8 (229 \tau -52)\\
&\quad-\frac{2}{9} \alpha ^6 (302 \tau -39)+\frac{4}{3} \alpha ^4 (22 \tau -3)\ , \\
 f^{B,v}_{de32} &= \frac{4}{3} \alpha ^{10} (4 \tau -1)-\frac{16}{9} \alpha ^8 (26 \tau -5)\\
&\quad+\frac{8}{3} \alpha ^6 (46 \tau -5)-48 \alpha ^4 (2 \tau -1)\ , \\
 f^{B,v}_{de33} &= \frac{4}{3} \alpha ^9 (4 \tau -1)-\frac{8}{9} \alpha ^7 (40 \tau -7)\\
&\quad+\frac{16}{9} \alpha ^5 (35 \tau -2)-\frac{64 \alpha ^3 \tau }{3}\ , \\
 f^{B,v}_{de34} &= \frac{2}{3} \alpha ^{10} (4 \tau -1)-\frac{8}{9} \alpha ^8 (26 \tau -5)\\
&\quad+\frac{4}{3} \alpha ^6 (46 \tau -5)-24 \alpha ^4 (2 \tau -1)\ , \\
 f^{B,v}_{de41} &= -\frac{5 \alpha ^8}{9}+\frac{44 \alpha ^6}{9}-\frac{35 \alpha ^4}{3}\ , \\
 f^{B,v}_{de42} &= \frac{2 \alpha ^8}{3}-\frac{20 \alpha ^6}{3}+20 \alpha ^4\ , \\
 f^{B,v}_{de43} &= \frac{2 \alpha ^7}{3}-\frac{16 \alpha ^5}{3}+\frac{32 \alpha ^3}{3}\ , \\
 f^{B,v}_{de44} &= \frac{\alpha ^8}{3}-\frac{10 \alpha ^6}{3}+10 \alpha ^4\ , \\
 f^{B,v}_{de51} &= -\frac{14 \alpha ^8}{9}+\frac{32 \alpha ^6}{3}-16 \alpha ^4\ , \\
 f^{B,v}_{de52} &= \frac{8 \alpha ^8}{3}-24 \alpha ^6+64 \alpha ^4\ , \\
 f^{B,v}_{de53} &= \frac{8 \alpha ^7}{3}-\frac{56 \alpha ^5}{3}+32 \alpha ^3\ , \\
 f^{B,v}_{de54} &= \frac{4 \alpha ^8}{3}-12 \alpha ^6+32 \alpha ^4\ , \\
 f^{B,v}_{de61} &= -\frac{8 \alpha ^8}{3}+\frac{128 \alpha ^6}{9}-\frac{38 \alpha ^4}{3}\ , \\
 f^{B,v}_{de62} &= 4 \alpha ^8-\frac{80 \alpha ^6}{3}+40 \alpha ^4\ , \\
 f^{B,v}_{de63} &= 4 \alpha ^7-\frac{56 \alpha ^5}{3}+\frac{32 \alpha ^3}{3}\ , \\
 f^{B,v}_{de64} &= 2 \alpha ^8-\frac{40 \alpha ^6}{3}+20 \alpha ^4\ .
\end{align*}
\subsection{\texorpdfstring{$C^v_{2}$}{C2v}}
\begin{subequations}
\begin{align}
C^v_{a} &= -2 l_{2,1} m_N \ ,\\
C^v_{b} &= 0 \ , \\
C^v_{c} &= \frac{g_A^2 m_N^2}{16 \pi^2 F_\pi^2 (1-\tilde{\tau})^4}\bigl(f^{C,v}_{c1} l_{0,1}+f^{C,v}_{c2} l_{1,2} m_N \bigr) \ ,\\
C^v_{de} &= \frac{g_A m_N^2}{16 \pi^2 F_\pi^2}\bigl(
\begin{aligned}[t]
&f^{C,v}_{de1} l_{1,1} m_N+f^{C,v}_{de2} l_{1,3} m_N+f^{C,v}_{de3} l_{1,8} m_N^3\\
&+f^{C,v}_{de4} (l_{1,13}+l_{1,14}) m_N^2\\
&+f^{C,v}_{de5} (l_{1,15}+l_{1,16}) m_N^2\\
&+f^{C,v}_{de6} (l_{1,18}+l_{1,19}) m_N\bigr) \ ,
\end{aligned} \raisetag{12pt}
\end{align}
\end{subequations}
\begin{align*}
f^{C,v}_{c11} &= \frac{10 \alpha ^6 u^2}{3}+4 \alpha ^4 (\tilde{\tau}-1) u^2\ , \\
f^{C,v}_{c12} &= -8 \alpha ^6 u^2-16 \alpha ^4 (\tilde{\tau}-1) u^2\ , \\
f^{C,v}_{c13} &= 0\ , \\
f^{C,v}_{c14} &= 8 \alpha ^7 u^2+32 \alpha ^5 (\tilde{\tau}-1) u^2+16 \alpha ^3 (\tilde{\tau}-1)^2 u^2\ , \\
f^{C,v}_{c15} &= -4 \alpha ^6 u^2-8 \alpha ^4 (\tilde{\tau}-1) u^2\ , \\
f^{C,v}_{c21} &= -\frac{7}{18} \alpha ^6 (\tilde{\tau}-1)-\frac{2}{3} \alpha ^4 (\tilde{\tau}-1)^2\ , \\
f^{C,v}_{c22} &= \frac{2}{3} \alpha ^6 (\tilde{\tau}-1)+2 \alpha ^4 (\tilde{\tau}-1)^2\ , \\
f^{C,v}_{c23} &= 0\ , \\
f^{C,v}_{c24} &= -\frac{2}{3} \alpha ^7 (\tilde{\tau}-1)-\frac{10}{3} \alpha ^5 (\tilde{\tau}-1)^2-\frac{8}{3} \alpha ^3 (\tilde{\tau}-1)^3\ , \\
f^{C,v}_{c25} &= \frac{1}{3} \alpha ^6 (\tilde{\tau}-1)+\alpha ^4 (\tilde{\tau}-1)^2\ , \\
 f^{C,v}_{de11} &= \frac{\alpha ^8}{6}-\frac{8 \alpha ^6}{9}+\frac{19 \alpha ^4}{24}\ , \\
 f^{C,v}_{de12} &= -\frac{\alpha ^8}{4}+\frac{5 \alpha ^6}{3}-\frac{5 \alpha ^4}{2}\ , \\
 f^{C,v}_{de13} &= -\frac{\alpha ^7}{4}+\frac{7 \alpha ^5}{6}-\frac{2 \alpha ^3}{3}\ , \\
 f^{C,v}_{de14} &= -\frac{\alpha ^8}{8}+\frac{5 \alpha ^6}{6}-\frac{5 \alpha ^4}{4}\ , \\
 f^{C,v}_{de21} &= -\frac{\alpha ^8}{3}+\alpha ^6-\frac{\alpha ^4}{4}\ , \\
 f^{C,v}_{de22} &= \frac{\alpha ^8}{2}-2 \alpha ^6+\alpha ^4\ , \\
 f^{C,v}_{de23} &= \frac{\alpha ^7}{2}-\alpha ^5\ , \\
 f^{C,v}_{de24} &= \frac{\alpha ^8}{4}-\alpha ^6+\frac{\alpha ^4}{2}\ , \\
 f^{C,v}_{de31} &= \frac{1}{6} \alpha ^{10} (7 \tau -4)+\frac{1}{27} \alpha ^8 (104-157 \tau )\\
&\quad+\frac{1}{3} \alpha ^6 (17 \tau -13)\ , \\
 f^{C,v}_{de32} &= -\frac{2}{3} \alpha ^{10} (2 \tau -1)+\frac{4}{9} \alpha ^8 (17 \tau -10)-\frac{4}{3} \alpha ^6 (7 \tau -5)\ , \\
 f^{C,v}_{de33} &= -\frac{2}{3} \alpha ^9 (2 \tau -1)+\frac{4}{9} \alpha ^7 (11 \tau -7)-\frac{4}{9} \alpha ^5 (5 \tau -4)\ , \\
 f^{C,v}_{de34} &= \frac{1}{3} \alpha ^{10} (1-2 \tau )+\frac{2}{9} \alpha ^8 (17 \tau -10)-\frac{2}{3} \alpha ^6 (7 \tau -5)\ , \\
 f^{C,v}_{de41} &= -\frac{5 \alpha ^8}{36}+\frac{2 \alpha ^6}{3}-\frac{\alpha ^4}{4}\ , \\
 f^{C,v}_{de42} &= \frac{\alpha ^8}{6}-\alpha ^6+\alpha ^4\ , \\
 f^{C,v}_{de43} &= \frac{\alpha ^7}{6}-\frac{2 \alpha ^5}{3}\ , \\
 f^{C,v}_{de44} &= \frac{\alpha ^8}{12}-\frac{\alpha ^6}{2}+\frac{\alpha ^4}{2}\ , \\
 f^{C,v}_{de51} &= \frac{7 \alpha ^8}{18}-\frac{2 \alpha ^6}{3}\ , \\
 f^{C,v}_{de52} &= 2 \alpha ^6-\frac{2 \alpha ^8}{3}\ , \\
 f^{C,v}_{de53} &= \frac{2 \alpha ^5}{3}-\frac{2 \alpha ^7}{3}\ , \\
 f^{C,v}_{de54} &= \alpha ^6-\frac{\alpha ^8}{3}\ , \\
 f^{C,v}_{de61} &= \frac{2 \alpha ^8}{3}-2 \alpha ^6+\frac{\alpha ^4}{2}\ , \\
 f^{C,v}_{de62} &= -\alpha ^8+4 \alpha ^6-2 \alpha ^4\ , \\
 f^{C,v}_{de63} &= 2 \alpha ^5-\alpha ^7\ , \\
 f^{C,v}_{de64} &= -\frac{\alpha ^8}{2}+2 \alpha ^6-\alpha ^4\ .
\end{align*}
\subsection{\texorpdfstring{$\tilde{A}^v_{20}$}{At20v}}
\begin{subequations}
\begin{align}
\tilde{A}^v_{a} &= 16 \alpha ^2 l_{2,4} m_N^2-16 \tau  l_{2,5} m_N^2+4 Z l_{0,2} \ ,\\
\tilde{A}^v_{b} &= -\frac{\alpha ^2 m_N^2 l_{0,2} }{4 \pi^2 F_\pi^2}f^v_b \ ,\\
\tilde{A}^v_{c} &= \frac{g_A^2 m_N^2}{16 \pi^2 F_\pi^2 (1-\tilde{\tau})^4}f^{\tilde{A},v}_{c1} l_{0,2} \ ,\\
\tilde{A}^v_{de} &= \frac{g_A m_N^2}{16 \pi^2 F_\pi^2}\bigl(
\begin{aligned}[t]
&f^{\tilde{A},v}_{de1} l_{0,1}+f^{\tilde{A},v}_{de2} l_{1,2} m_N+f^{\tilde{A},v}_{de3} l_{1,17} m_N\\
&+f^{\tilde{A},v}_{de4} (l_{1,4}+l_{1,5})+f^{\tilde{A},v}_{de5} (l_{1,9}+l_{1,10}) m_N^2\\
&+f^{\tilde{A},v}_{de6} (l_{1,11}+ l_{1,12}) m_N^2\bigr) \ ,
\end{aligned} \raisetag{12pt}
\end{align}
\end{subequations}
\begin{align*}
f^{\tilde{A},v}_{c11} &= -\frac{7}{9} \alpha ^6 (\tilde{\tau}-1) \left(\tilde{\tau}^3-3 \tilde{\tau}^2+3 \tilde{\tau}+1\right)\\
&\quad+\frac{1}{3} \alpha ^4 (\tilde{\tau}-1)^2 \left(7 \tilde{\tau}^2-14 \tilde{\tau}-4\right)-2 \alpha ^2 (\tilde{\tau}-1)^3\ , \\
f^{\tilde{A},v}_{c12} &= \frac{4}{3} \alpha ^6 (\tilde{\tau}-1) \left(\tilde{\tau}^3-3 \tilde{\tau}^2+3 \tilde{\tau}+1\right)\\
&\quad-6 \alpha ^4 (\tilde{\tau}-2) (\tilde{\tau}-1)^2 \tilde{\tau}+2 \alpha ^2 (\tilde{\tau}-1)^4\ , \\
f^{\tilde{A},v}_{c13} &= \frac{4}{3} \alpha ^5 (\tilde{\tau}-1)^4-\frac{10}{3} \alpha ^3 (\tilde{\tau}-1)^4\ , \\
f^{\tilde{A},v}_{c14} &= -\frac{8}{3} \alpha ^7 (\tilde{\tau}-1)-\frac{34}{3} \alpha ^5 (\tilde{\tau}-1)^2-\frac{20}{3} \alpha ^3 (\tilde{\tau}-1)^3\ , \\
f^{\tilde{A},v}_{c15} &= \frac{2}{3} \alpha ^6 (\tilde{\tau}-1) \left(\tilde{\tau}^3-3 \tilde{\tau}^2+3 \tilde{\tau}+1\right)\\
&\quad-3 \alpha ^4 (\tilde{\tau}-2) (\tilde{\tau}-1)^2 \tilde{\tau}+\alpha ^2 (\tilde{\tau}-1)^4\ , \\
 f^{\tilde{A},v}_{de11} &= \frac{7 \alpha ^6}{9}-\frac{10 \alpha ^4}{3}\ , \\
 f^{\tilde{A},v}_{de12} &= 8 \alpha ^4-\frac{4 \alpha ^6}{3}\ , \\
 f^{\tilde{A},v}_{de13} &= \frac{16 \alpha ^3}{3}-\frac{4 \alpha ^5}{3}\ , \\
 f^{\tilde{A},v}_{de14} &= 4 \alpha ^4-\frac{2 \alpha ^6}{3}\ , \\
 f^{\tilde{A},v}_{de21} &= \frac{5 \alpha ^8}{18}-\frac{8 \alpha ^6}{9}-\frac{5 \alpha ^4}{6}\ , \\
 f^{\tilde{A},v}_{de22} &= -\frac{\alpha ^8}{3}+\frac{2 \alpha ^6}{3}+6 \alpha ^4\ , \\
 f^{\tilde{A},v}_{de23} &= \frac{16 \alpha ^3}{3}-\frac{\alpha ^7}{3}\ , \\
 f^{\tilde{A},v}_{de24} &= -\frac{\alpha ^8}{6}+\frac{\alpha ^6}{3}+3 \alpha ^4\ , \\
 f^{\tilde{A},v}_{de31} &= \frac{5 \alpha ^8}{9}-\frac{16 \alpha ^6}{9}-\frac{5 \alpha ^4}{3}\ , \\
 f^{\tilde{A},v}_{de32} &= -\frac{2 \alpha ^8}{3}+\frac{4 \alpha ^6}{3}+12 \alpha ^4\ , \\
 f^{\tilde{A},v}_{de33} &= \frac{32 \alpha ^3}{3}-\frac{2 \alpha ^7}{3}\ , \\
 f^{\tilde{A},v}_{de34} &= -\frac{\alpha ^8}{3}+\frac{2 \alpha ^6}{3}+6 \alpha ^4\ , \\
 f^{\tilde{A},v}_{de41} &= \frac{8 \alpha ^4}{3}-\frac{14 \alpha ^6}{9}\ , \\
 f^{\tilde{A},v}_{de42} &= \frac{8 \alpha ^6}{3}-8 \alpha ^4\ , \\
 f^{\tilde{A},v}_{de43} &= \frac{8 \alpha ^5}{3}-\frac{8 \alpha ^3}{3}\ , \\
 f^{\tilde{A},v}_{de44} &= \frac{4 \alpha ^6}{3}-4 \alpha ^4\ , \\
 f^{\tilde{A},v}_{de51} &= \frac{8 \alpha ^8 \tau }{3}-\frac{128 \alpha ^6 \tau }{9}+\frac{1}{3} \alpha ^4 (38 \tau -3)\ , \\
 f^{\tilde{A},v}_{de52} &= -4 \alpha ^8 \tau +\frac{80 \alpha ^6 \tau }{3}-20 \alpha ^4 (2 \tau -1)\ , \\
 f^{\tilde{A},v}_{de53} &= -4 \alpha ^7 \tau +\frac{56 \alpha ^5 \tau }{3}-\frac{32 \alpha ^3 \tau }{3}\ , \\
 f^{\tilde{A},v}_{de54} &= -2 \alpha ^8 \tau +\frac{40 \alpha ^6 \tau }{3}-10 \alpha ^4 (2 \tau -1)\ , \\
 f^{\tilde{A},v}_{de61} &= -2 \alpha ^4\ , \\
 f^{\tilde{A},v}_{de62} &= 8 \alpha ^4\ , \\
 f^{\tilde{A},v}_{de63} &= 0\ , \\
 f^{\tilde{A},v}_{de64} &= 4 \alpha ^4\ .
\end{align*}
\subsection{\texorpdfstring{$\tilde{B}^v_{20}$}{Bt20v}}
\begin{subequations}
\begin{align}
\tilde{B}^v_{a} &= 32 \alpha ^2 l_{3,1} m_N^3-32 \tau  l_{3,2} m_N^3+8 Z l_{1,1} m_N \ ,\\
\tilde{B}^v_{b} &= -\frac{\alpha ^2 m_N^2 l_{1,1} m_N }{2 \pi^2 F_\pi^2}f^v_b \ ,\\
\tilde{B}^v_{c} &= \frac{g_A^2 m_N^2}{16 \pi^2 F_\pi^2 (1-\tilde{\tau})^4}\big(f^{\tilde{B},v}_{c1} l_{0,2}+f^{\tilde{B},v}_{c2} l_{1,1} m_N\big) \ ,\\
\tilde{B}^v_{de} &= \frac{g_A m_N^2}{16 \pi^2 F_\pi^2}\big(
\begin{aligned}[t]
&f^{\tilde{B},v}_{de1} l_{1,2} m_N+f^{\tilde{B},v}_{de2} l_{1,17} m_N\\
&+f^{\tilde{B},v}_{de3} (l_{1,9}+l_{1,10}) m_N^2\\
&+f^{\tilde{B},v}_{de4} (l_{1,11}+ l_{1,12}) m_N^2\big) \ ,
\end{aligned}
\end{align}
\end{subequations}
\begin{align*}
f^{\tilde{B},v}_{c11} &= -\frac{80}{3} \alpha ^6 u^2-40 \alpha ^4 (\tilde{\tau}-1) u^2\ , \\
f^{\tilde{B},v}_{c12} &= 64 \alpha ^6 u^2+160 \alpha ^4 (\tilde{\tau}-1) u^2+32 \alpha ^2 (\tilde{\tau}-1)^2 u^2\ , \\
f^{\tilde{B},v}_{c13} &= 0\ , \\
f^{\tilde{B},v}_{c14} &= -64 \alpha ^7 u^2-288 \alpha ^5 (\tilde{\tau}-1) u^2-224 \alpha ^3 (\tilde{\tau}-1)^2 u^2\ , \\
f^{\tilde{B},v}_{c15} &= 32 \alpha ^6 u^2+80 \alpha ^4 (\tilde{\tau}-1) u^2+16 \alpha ^2 (\tilde{\tau}-1)^2 u^2\ , \\
f^{\tilde{B},v}_{c21} &= \frac{14}{9} \alpha ^6 (\tilde{\tau}-1)^4-\frac{14}{3} \alpha ^4 (\tilde{\tau}-1)^4-4 \alpha ^2 (\tilde{\tau}-1)^3\ , \\
f^{\tilde{B},v}_{c22} &= -\frac{8}{3} \alpha ^6 (\tilde{\tau}-1)^4+4 \alpha ^4 (\tilde{\tau}-1)^2 \left(3 \tilde{\tau}^2-6 \tilde{\tau}+1\right)\\
&\quad-4 \alpha ^2 (\tilde{\tau}-1)^3 (\tilde{\tau}+1)\ , \\
f^{\tilde{B},v}_{c23} &= \frac{20}{3} \alpha ^3 (\tilde{\tau}-1)^4-\frac{8}{3} \alpha ^5 (\tilde{\tau}-1)^4\ , \\
f^{\tilde{B},v}_{c24} &= 8 \alpha ^5 (\tilde{\tau}-1)^2+24 \alpha ^3 (\tilde{\tau}-1)^3\ , \\
f^{\tilde{B},v}_{c25} &= -\frac{4}{3} \alpha ^6 (\tilde{\tau}-1)^4+2 \alpha ^4 (\tilde{\tau}-1)^2 \left(3 \tilde{\tau}^2-6 \tilde{\tau}+1\right)\\
&\quad-2 \alpha ^2 (\tilde{\tau}-1)^3 (\tilde{\tau}+1)\ , \\
 f^{\tilde{B},v}_{de11} &= -\frac{4 \alpha ^8}{3}+\frac{64 \alpha ^6}{9}-\frac{19 \alpha ^4}{3}\ , \\
 f^{\tilde{B},v}_{de12} &= 2 \alpha ^8-\frac{40 \alpha ^6}{3}+20 \alpha ^4\ , \\
 f^{\tilde{B},v}_{de13} &= 2 \alpha ^7-\frac{28 \alpha ^5}{3}+\frac{16 \alpha ^3}{3}\ , \\
 f^{\tilde{B},v}_{de14} &= \alpha ^8-\frac{20 \alpha ^6}{3}+10 \alpha ^4\ , \\
 f^{\tilde{B},v}_{de21} &= -\frac{8 \alpha ^8}{3}+\frac{100 \alpha ^6}{9}-\frac{22 \alpha ^4}{3}\ , \\
 f^{\tilde{B},v}_{de22} &= 4 \alpha ^8-\frac{64 \alpha ^6}{3}+24 \alpha ^4\ , \\
 f^{\tilde{B},v}_{de23} &= 4 \alpha ^7-\frac{40 \alpha ^5}{3}+\frac{16 \alpha ^3}{3}\ , \\
 f^{\tilde{B},v}_{de24} &= 2 \alpha ^8-\frac{32 \alpha ^6}{3}+12 \alpha ^4\ , \\
 f^{\tilde{B},v}_{de31} &= \frac{20 \alpha ^6}{9}-\frac{32 \alpha ^4}{3}\ , \\
 f^{\tilde{B},v}_{de32} &= 16 \alpha ^4-\frac{8 \alpha ^6}{3}\ , \\
 f^{\tilde{B},v}_{de33} &= \frac{32 \alpha ^3}{3}-\frac{8 \alpha ^5}{3}\ , \\
 f^{\tilde{B},v}_{de34} &= 8 \alpha ^4-\frac{4 \alpha ^6}{3}\ , \\
 f^{\tilde{B},v}_{de41} &= \frac{28 \alpha ^8}{9}-\frac{40 \alpha ^6}{3}\ , \\
 f^{\tilde{B},v}_{de42} &= -\frac{16 \alpha ^8}{3}+32 \alpha ^6-32 \alpha ^4\ , \\
 f^{\tilde{B},v}_{de43} &= \frac{64 \alpha ^5}{3}-\frac{16 \alpha ^7}{3}\ , \\
 f^{\tilde{B},v}_{de44} &= -\frac{8 \alpha ^8}{3}+16 \alpha ^6-16 \alpha ^4\ .
\end{align*}

%
%

\bibliographystyle{apsrev}
\providecommand{\href}[2]{#2}\begingroup\raggedright\endgroup

\end{document}